\documentclass[iop]{emulateapj}
\usepackage{natbib}
\usepackage{graphicx}
\usepackage{color}
\usepackage{amsmath, amssymb}

\shorttitle{Mass-to-Light Variations in 3D FP Space}
\shortauthors{Graves \& Faber}

\begin{document}

\title{Dissecting the Red Sequence---III. Mass-to-Light Variations 
\\ in 3D Fundamental Plane Space}

\author{Genevieve J. Graves\altaffilmark{1,2,3} \&
  S. M. Faber\altaffilmark{1}}

\altaffiltext{1}{UCO/Lick Observatory, Department of Astronomy and
  Astrophysics, University of California, Santa Cruz, CA 95064, USA}
\altaffiltext{2}{Department of Astronomy, University of California,
  Berkeley, CA 94720, USA; graves@astro.berkeley.edu}
\altaffiltext{3}{Miller Fellow}


\begin{abstract}
The Fundamental Plane of early type galaxies is observed to have
finite thickness and to be tilted from the virial relation.  Both of
these represent departures from the simple assumption that dynamical
mass-to-light ratios ($M_{dyn}/L$) are constant for all early type
galaxies.  We use a sample of 16,000 quiescent galaxies from the Sloan
Digital Sky Survey to map out the variations in $M_{dyn}/L$ throughout
the 3D Fundamental Plane space defined by velocity dispersion
($\sigma$), effective radius ($R_e$), and effective surface brightness
($I_e$).  Dividing $M_{dyn}/L$ into multiple components allows us to
separately consider the contribution to the observed $M_{dyn}/L$
variation due to stellar population effects, IMF variations, and
variations in the dark matter fraction within one $R_e$.  Along the
FP, we find that the stellar population contribution given some
constant IMF ($M_{\star,IMF}/L$) scales with $\sigma$ such that
$M_{\star,IMF}/L \propto f(\sigma)$.  Meanwhile, the dark matter
and/or IMF contribution ($M_{dyn}/M_{\star,IMF}$) scales with
$M_{dyn}$ such that $M_{dyn}/M_{\star,IMF} \propto g(M_{dyn})$.  This
means that the two contributions to the tilt of the FP rotate the
plane around different axes in the 3D space.  The observed tilt of the
FP requires contributions from both, with dark matter/IMF variations
likely comprising the dominant contribution.  Looking at $M_{dyn}/L$
variations through the {\it thickness} of the FP, we find that
$M_{dyn}/L$ variations must be dominated either by IMF variations or
by real differences in dark matter fraction with $R_e$.  This means
that the finite thickness of the FP is due to variations in the
stellar mass surface density within $R_e$ ($\Sigma_{\star,IMF}$), not
the fading of passive stellar populations and it therefore represents
genuine structural differences between early type galaxies.  These
structural variations are correlated with galaxy star formation
histories such that galaxies with higher $M_{dyn}/M_{\star,IMF}$ have
higher [Mg/Fe], lower metallicities, and older mean stellar ages.  We
discuss several physical mechanisms that might explain the observed
co-variation between $M_{dyn}/M_{\star,IMF}$ and galaxy star formation
histories.  It is difficult to explain the observed enhancement of
$\alpha$-elements in lower-surface-brightness galaxies by allowing the
IMF to vary.  Differences in dark matter fraction can be produced by
variations in the ``conversion efficiency'' of baryons into stars or
by the redistribution of stars and dark matter through dissipational
merging.  The former explanation, specifically a model in which some
galaxies experience low conversion efficiencies due to premature
truncation of star formation, provides a more natural explanation for
the co-variation of $M_{dyn}/M_{\star,IMF}$ and the observed stellar
population properties.
\end{abstract}
 
\section{Introduction}\label{introduction}

Early type galaxies are observed to obey many scaling relations among
their structural properties.  Early work identified a number of 1D
relations, such as the Faber-Jackson relation between galaxy
luminosity ($L$) and velocity dispersion ($\sigma$) \citep{faber76},
variations in galaxy mass-to-light ratio ($M/L$) versus $L$
(\citealt{tinsley81}; \citealt{faber87}), and various correlations of
$\sigma$ and $L$ with galaxy effective radius ($R_e$) and effective
surface brightness ($\mu_e$) \citep{kormendy85}, or with galaxy core
radius ($r_c$) and central surface brightness ($I_{0}$)
\citep{lauer85}.  

These relations generally reflect a 1D mass sequence of galaxies.  It
is clear however that early type galaxies comprise at least a
two-parameter family in terms of their structure.  In the
three-dimensional parameter space of galaxy properties defined by
$\sigma$, $R_e$, and surface brightness (expressed as $\mu_e$ in
magnitudes or as $I_e$ in linear flux units), early type galaxies
populate a relatively tight two-dimensional plane, known as the
Fundamental Plane (FP, \citealt{djorgovski87, dressler87}).
Projections of the FP appear narrow in some orientations, leading to
the seemingly 1D relations listed above.

The FP can be understood as a manifestation of the virial plane
predicted for relaxed systems, under the assumption that galaxy
mass-to-light ratios ($M_{dyn}/L$) are constant (or at least smoothly
varying) for all galaxies.  If $M_{dyn}/L$ were strictly constant and
if structures were homologous for all early type galaxies, the FP
would be equivalent to the virial plane, which takes the form $R_e
\propto \sigma^2 I_e^{-1}$, and would be infinitely thin.  Instead,
the FP is rotated from the virial plane (e.g., \citealt{dressler87,
  djorgovski87}).  \citet{jorgensen96} find $R_e \propto \sigma^{1.24}
I_e^{-0.82}$ for the FP of local cluster early type galaxies, as
measured in the Gunn $r$ band.  In addition, the FP has finite
thickness (e.g., \citealt{jorgensen96, forbes98, wuyts04, hyde09b,
  gargiulo09}) indicating further departures from a single virial
plane.  The ``tilt'' of the FP and its finite thickness indicate
departures from the simplistic assumption that $M_{dyn}/L$ is
constant for all early type galaxies and that their structures are
homologous.

This is the third in a series of four papers that explores the mapping
between the 2D family of early type galaxy stellar populations and
their structural properties.  The first three papers in the series
systematically gather together a number of galaxy properties that vary
through the different dimensions of galaxy parameter space.
\citet[hereafter Paper I]{graves09_paperI} showed that stellar
populations vary differently as functions of $\sigma$ than as
functions of $L$.  This systematic variation suggests an underlying
multi-dimensional parameter space.  \citet[hereafter Paper
  II]{graves09_paperII} mapped out the second dimension of stellar
population properties in 3D Fundamental Plane space and correlated
stellar population variations with FP parameters.  In this paper, we
focus on the corresponding mass-to-light ratios and how they relate to
the distribution of galaxies in FP space.

We show here that the stellar population variations measured in Paper
II do not provide enough variation in galaxy mass-to-light ratios to
explain either the tilt or the thickness of the FP.  Some correlated
variation is also required in either the initial mass function (IMF)
with which galaxies form their stars, or in the dark matter fraction
inside $R_e$.  At the same time, these variations are a key component
of structural differentiation in galaxies that must also be related to
their star formation histories.  The results of these first three
papers are gathered together in Paper IV (Graves et al. 2010,
submitted to ApJ), which uses them to propose scenarios for galaxy
formation and evolution through the thickness of the FP.

Section \ref{data} briefly outlines the sample of galaxies used in
this work, as well as our methods for grouping together similar
galaxies to produce high signal-to-noise spectra that span the space
of galaxy properties.  In section \ref{ml_contributions}, we discuss
the various physical mechanisms that may contribute to the observed
variations in $M_{dyn}/L$ throughout the FP.  Some of this variation
is due to stellar population effects, but quantifying these depends
critically on the stellar population models used to measure stellar
mass-to-light ratios ($M_{\star,IMF}/L$).  We therefore investigate
multiple ways of measuring $M_{\star,IMF}/L$ from stellar population
models in section \ref{mstar_l}.  This comparison allows us to explore
the range of possible $M_{\star,IMF}/L$ values for our sample galaxies
and to understand the biases inherent in each model.  We demonstrate
that, regardless of which method is chosen to measure
$M_{\star,IMF}/L$, the observed stellar population variations are
inadequate to explain either the tilt or the thickness of the FP.
Section \ref{ml_in_3d} shows that the two contributions to the FP tilt
(1: from known stellar population variations and 2: from IMF and/or
dark matter fraction variations within $R_e$) rotate the plane around
different axes in 3D Fundamental Plane space.  It also investigates
the various contributions to the {\it thickness} of the FP and shows
that the spread in $M_{dyn}/L$ through the plane is due primarily to
the second factor: variations in the IMF and/or dark matter fractions.
Section \ref{sfh} discusses physical models that may produce the
observed correlations between dark matter fraction or IMF variation
and galaxy star formation histories.  It illustrates that the observed
trends can be explained by a model in which some galaxies experience
low ``conversion efficiency'' turning baryons into stars due to
trucated star formation histories.  Finally, section \ref{conclusions}
summarizes our conclusions.

\begin{figure*}[t]
\includegraphics[width=1.0\linewidth]{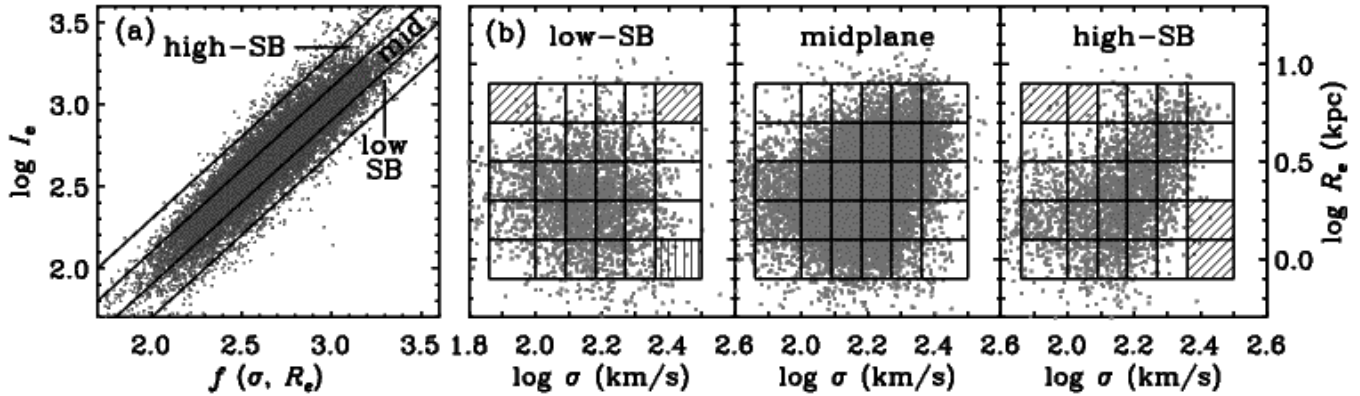}
\caption{Sorting and binning galaxies in 3D FP space.  (a)~An edge-on
  view of the best fit to the FP, using a least-squares fit of $I_e$
  onto $R_e$ and $\sigma$, which gives $f(\sigma,R_e) = 1.16 \log
  \sigma - 1.21 \log R_e + 0.55$.  We divide the FP into three slices
  based on surface brightness residuals ($\Delta \log I_e$) to define
  low-surface-brightness, midplane, and high-surface-brightness
  slices. (b)~Within each slice, galaxies are further divided into 6
  bins in $\log \sigma$ and 5 bins in $\log R_e$, as illustrated.
  Diagonal shading indicates the six bins that do not contain enough
  galaxies to produce robust mean spectra and are therefore excluded
  from our analysis.  Vertical shading indicates a bin where stellar
  population ages cannot be reliably measured (see Paper II).
}\label{fp_bins}
\end{figure*}

\section{Data}\label{data}

The data used in this analysis are the same as those used in Papers I
and II.  They consist of a sample of $\sim$16,000 non-star-forming
(quiescent) galaxies chosen from the SDSS spectroscopic Main Galaxy
Sample \citep{strauss02} in a relatively narrow redshift range.  The
sample selection is described in detail in Paper I.

Briefly, we construct a sample of quiescent galaxies by requiring that
their spectra contain no detectable emission in either H$\alpha$ or
the [O\textsc{ii}]$\lambda$3727 doublet (all measured line strengths
are below a $2 \sigma$ detection threshold)\footnote{This emission cut
  excludes known star-forming galaxies and also LINER hosts.  The
  LINER hosts can be considered quiescent systems, in the sense that
  they do not contain ongoing star formation, but they have
  systematically younger stellar populations \citep{graves07} and have
  larger corrections to H$\beta$ for emission infill.  We will return
  to LINERs in a future paper.}, based on the emission line
measurements of \citet{yan06} for SDSS Data Release 4 (DR4,
\citealt{adelman-mccarthy06}) galaxies.  We further limit the sample
to a relatively narrow redshift slice ($0.04 < z < 0.08$) and to
galaxies with light profiles consistent with early type morphologies.
These criteria produce a sample of galaxies that fall on the red
sequence in a color-magnitude diagram (see Figure 1 of Paper I), and
which typically have bulge-dominated morphologies.

The various FP parameters are obtained for all galaxies in the sample,
as follows.  Photometry and $R_e$ values are from de Vaucouleurs fits
to the light profile.  The radii and $\sigma$ values are downloaded
from the SDSS Catalog Archive
Server\footnote{http://cas.sdss.org/dr6/en/} and the apparent
photometry from the NYU Value-Added Galaxy Catalog
\citep{blanton05-vagc}.  The photometry is subsequently corrected for
Galactic extinction \citep{schlegel98}, K-corrected to the $V$-band at
$z=0$ using the IDL code {\it kcorrect} v4.1.4 \citep{blanton07}, and
converted to absolute luminosities assuming a standard $\Lambda$CDM
cosmology with $\Omega_{\Lambda}=0.7$, $\Omega_M=0.3$, and $h_0=0.7$.
Both $L$ and $R_e$ have been corrected for known problems with the
SDSS pipeline sky-subtraction around bright galaxies, as described in
Paper I.  The surface brightness $I_e$ is then computed as $I_e = L_V
/ 2 \pi R_e^2$.  All surface brightnesses and luminosities used in
this work are measured in the $V$-band.  The $\sigma$ values are
aperture-corrected to a constant $R_e/8$ aperture following
\citet{jorgensen95} and \citet{bernardi03a}.  Spectra for the sample
galaxies were downloaded from the SDSS Data Archive
Server\footnote{http://das.sdss.org/DR6-cgi-bin/DAS}.

The individual SDSS spectra of our galaxies typically have $S/N \sim
20$ \AA$^{-1}$, while accurate stellar population analysis requires
$S/N \ge 100$ \AA$^{-1}$ \citep{cardiel98}.  It is possible to make
estimates of galaxy stellar population properties from individual SDSS
spectra (e.g., \citealt{gallazzi05}) but we have chosen to take a
different approach by defining bins that group together ``similar''
galaxies, then stacking the spectra of all galaxies in each bin to
produce high $S/N$ mean spectra that track the average properties of
galaxies throughout the binning space.  The ideal $S/N > 100$
{\AA}$^{-1}$ requires $\sim 25$ galaxies per bin, depending on the
$S/N$ of the individual spectra that go into the composite.

In the analysis presented here, we use two different binning
strategies to explore mass-to-light ratio variations along the FP and
through its thickness.  The first binning scheme is the one used in
Paper II, which sorts and bins galaxies in the 3D parameter space
defined by $\sigma$, $R_e$, and $\Delta I_e$.\footnote{For brevity, we
  will use the terms $\sigma$, $R_e$ and $\Delta I_e$ throughout this
  paper to refer the the various dimensions of FP space.  It should be
  understood that all bins are defined in log space (i.e., $\log
  \sigma$, $\log R_e$, and $\Delta \log I_e$) and that we always
  measure each quantity in logarithmic units.}  This parameter space
is ideal for studying trends in galaxy properties along a (nearly)
face-on projection of the FP itself.  The edge-on projection of this
plane is shown in Figure \ref{fp_bins}a.  We divide FP space into
three slices based on residuals from the FP in the $I_e$ dimension
($\Delta \log I_e$).  These include a ``midplane'' slice, as well as
high- and low-surface-brightness slices above and below the plane.  We
then define a $6 \times 5$ grid in $\log \sigma$ and $\log R_e$ within
each slice (Figure \ref{fp_bins}b).  Bin widths in $\log \sigma$,
$\log R_e$ and $\log \Delta I_e$ (0.09 dex, 0.20 dex, and 0.10 dex,
respectively) are substantially larger than the uncertainties in the
measured quantities (typically $\sim 0.04$ dex, $\sim0.02$ dex, and
$\sim 0.03$ dex, respectively) so that the assignment of a galaxy to a
given bin is relatively robust.  We exclude bins that include fewer
than 10 galaxies and/or have composite spectra with $S/N < 50$
{\AA}$^{-1}$ (shaded regions in Figure \ref{fp_bins}b).

\begin{figure}[t]
\includegraphics[width=1.0\linewidth]{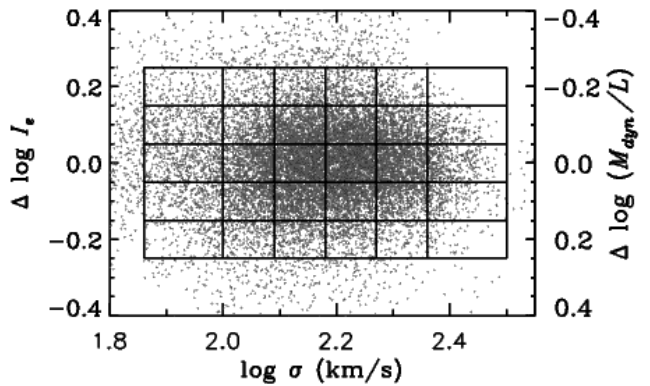}
\caption{Sorting and binning galaxies through the thickness of the FP.
  The quantity $\Delta \log I_e$ is defined as in Figure
  \ref{fp_bins}a.  We sort the sample into the same 6 bins in $\sigma$
  as in Figure \ref{fp_bins}b, and further into 5 bins in $\Delta \log
  I_e$.  }\label{fp_xsection_bins}
\end{figure}

In \S\ref{fp_thickness} we examine trends through the {\it thickness}
of the FP.  For this section, we bin galaxies in a 2D binning space
defined by $\sigma$ and $\Delta I_e$ only.  In Figure
\ref{fp_xsection_bins}, we show the distribution of the sample
galaxies in $\sigma$--$\Delta I_e$ space, which is essentially a
cross-section projected through the thickness of the FP.  We define a
$6 \times 5$ grid in this space and use it to sort galaxies into bins
through the FP cross-section.  Bins in $\sigma$ are chosen to be
consistent with the 3D binning scheme of Figure \ref{fp_bins}.  Bins
in $\Delta I_e$ span the majority of the data; the bin width is chosen
such that at high $\sigma$, even the outlying $\Delta I_e$ bins
contain $> 20$ galaxies and produce stacked spectra with $S/N > 100$
{\AA}$^{-1}$.  This binning scheme means that galaxies with a variety
of $R_e$ values are combined together in each bin.  However, as shown
in Paper II, the stellar population properties and therefore the star
formation histories of quiescent galaxies do not depend on $R_e$ at
fixed $\sigma$.  

With $\Delta I_e$ defined in this way, there is a direct
correspondence between $I_e$ variations and variations in $M_{dyn}/L$.
Recall that $M_{dyn}/L \propto \sigma^2 R_e / I_e R_e^2 \propto
\sigma^2 / I_e R_e$.  Variations in $I_e$ are defined at fixed points
in $\sigma$ and $R_e$.  Thus variations in $I_e$ at fixed values of
$\sigma$ and $R_e$ correspond to variations in $M_{dyn}/L$, such that
$\Delta I_e \propto \Delta (M_{dyn}/L)^{-1}$.

Once the galaxies have been sorted into bins, the spectra of all
galaxies in each bin are stacked together using an algorithm that
masks areas around bright sky lines and rejects highly deviant pixels.
We measure the full set of Lick indices in the resulting mean spectra,
then use them to model the mean stellar population properties of the
galaxies in each bin using the single burst population models of
\citet{schiavon07} and the code EZ\_Ages, described in
\citet{graves08}.  Sections 4.2--4.4 of Paper I give a detailed
description of this process.

There is one significant difference between the data presented here
and those of Paper II: here the stacked galaxy spectra have been
corrected for small quantities of emission infill in the H$\beta$
absorption line used to determine ages, as described in Appendix
\ref{oIII}.  The quantity of emission infill is similar in the various
stacked spectra.  Therefore the main effect of this correction is to
systematically lower the derived ages by $\sim 0.12$ dex and to raise
[Fe/H] values by $\sim 0.06$ dex compared to Paper II.  

\begin{figure}[t]
\includegraphics[width=1.0\linewidth]{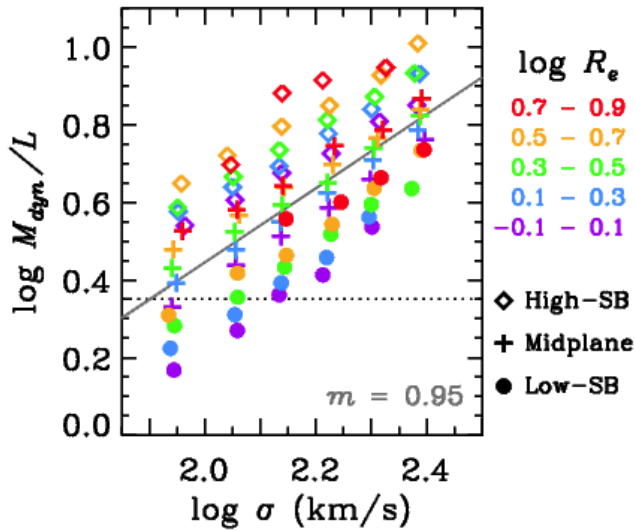}
\caption{Median values of $\sigma$ and $M_{dyn}/L$ for each bin of
  galaxies (3D FP binning).  The listed values of $R_e$ are in kpc.
  The solid line shows a least-squares fit of $M_{dyn}/L$ onto
  $\sigma$ for galaxies on the FP midplane, with the slope of the
  relation indicated in the lower right corner.  If the FP and the
  virial plane were equivalent, $M_{dyn}/L$ would be constant for all
  galaxies.  Instead, there is a strong increase in $M_{dyn}/L$ as
  $\sigma$ increases (the ``$\sigma$ tilt'' of the FP) and a
  substantial spread in $M_{dyn}/L$ at fixed $\sigma$ and $R_e$ (the
  ``thickness'' of the FP).  }\label{vdisp_mdyn}
\end{figure}

\section{Physical Effects that Contribute to $M_{dyn}/L$
  Variations Throughout the Fundamental Plane}\label{ml_contributions}

As discussed above, both the tilt of the FP and its finite thickness
represent departures from the simple assumption of constant
$M_{dyn}/L$.  This is illustrated in Figure \ref{vdisp_mdyn} for the
galaxies in our sample, which span almost an order of magnitude in
$M_{dyn}/L$.  The systematic variation of $M_{dyn}/L$ with $\sigma$ is
a 2D projection of the 3D ``tilt'' of the FP.  This $\sigma$-component
to the FP tilt has a slope of $M_{dyn}/L \propto \sigma^{0.95}$ in the
data presented here, based on a least squares fit that weights each
binned data point by the number of galaxies in the bin.  The tilt of
the FP is typically parameterized as a function of $M_{dyn}$, but it
is sometimes useful to discuss the $\sigma$ component of the tilt
separately (as in \S\ref{fp_tilt}, where we show that the stellar
population contribution to the tilt is purely a function of $\sigma$).
Throughout this paper, we will use the term ``$\sigma$ tilt'' to
discuss mass-to-light variation along the FP as a function of $\sigma$
and the term ``total tilt'' to describe variation along the FP as a
function of $M_{dyn}$.

In addition to the $\sigma$ tilt shown in Figure \ref{vdisp_mdyn},
$M_{dyn}/L$ also varies significantly at fixed $\sigma$.  This
variation is dominated by differences in $\Delta I_e$ through the
thickness of the FP (different symbols in Figure \ref{vdisp_mdyn}),
although $M_{dyn}/L$ also varies with $R_e$ at fixed $\sigma$
(different colors in Figure \ref{vdisp_mdyn}).

The FP appears to be the same in all environments
(\citealt{kochanek00, de_la_rosa01, de_carvalho03, reda05}, but see
\citealt{bernardi03c}) and includes both elliptical and S0 galaxies on
the same plane \citep{jorgensen96,fritz05}.  The slope of the FP
evolves with redshift, consistent with the passive aging of the
stellar populations (e.g., \citealt{kelson97, jorgensen99b,
  bernardi03c, van_de_ven03, holden05, jorgensen06, van_der_marel07b})
or with passive evolution plus a small quantity of residual star
formation \citep{gebhardt03}, but the thickness of the FP is constant
out to $z \sim 0.6$ \citep{kelson97, jorgensen99b, treu01}.  Thus the
tilt of the FP appears to be time-dependent but the thickness of the
FP does not, suggesting that these two departures from constant
$M_{dyn}/L$ may be driven by different mechanisms.

A number of different physical processes can contribute to the
variations in $M_{dyn}/L$ that produce the tilt and thickness of the
FP.  These include differences in the dynamical structure of galaxies,
variations in the ratio of dark matter to stellar mass, variations in
the intial mass function (IMF) with which galaxies form their stars,
and difference in the stellar mass-to-light ratios due to the star
formation rates, ages and metallicities of the galaxies.

To elucidate the contributions of each of these physical processes, we
separate the quantity $M_{dyn}/L$ into four components as follows:
\begin{equation}\label{ml_equation4}
\frac{M_{dyn}}{L} = \frac{M_{dyn}}{M_{tot}} \times
\frac{M_{tot}}{M_{\star}} \times \frac{M_{\star}}{M_{\star,IMF}} \times 
\frac{M_{\star,IMF}}{L},
\end{equation}
where $M_{dyn}$ is the dynamical mass estimate for the galaxy given by
$M_{dyn} \equiv 5 \sigma^2 R_e / G$ (e.g., \citealt{cappellari06}),
$M_{tot}$ is the total mass of the galaxy inside $R_e$, $M_{\star}$ is
the true {\it stellar} mass of the galaxy inside $R_e$,
$M_{\star,IMF}$ is the estimated stellar mass inside $R_e$ assuming
some constant IMF, and $L$ is the $V$-band luminosity inside $R_e$
(all quantities include $L$ are projected within $R_e$).  With these
definitions, the components of equation \ref{ml_equation4} represent
the following physical properties:
\begin{list}{}{}
\item[$\bullet$] $M_{dyn}/M_{tot}$ is the ``dynamical structure
  term.''  This accounts for any discrepancies between the simple
  dynamical mass estimator and the total projected mass within $R_e$.
\item[$\bullet$] $M_{tot}/M_{\star}$ is the ``dark matter term.''  It
  accounts for differences between the total projected mass within
  $R_e$ and the projected luminous stellar mass within $R_e$.
\item[$\bullet$] $M_{\star}/M_{\star,IMF}$ is the ``IMF term.''  It
  accounts for differences between the true projected stellar mass
  inside $R_e$ and the projected stellar mass computed with an assumed
  initial mass function.  If the true IMF matches the IMF used in the
  stellar population modelling, $M_{\star}/M_{\star,IMF} = 1$.
\item[$\bullet$] $M_{{\star},IMF}/L$ is the ``stellar population
  term.''  It accounts for the stellar mass-to-light ratio, given an
  assumed IMF.  The stellar population term can be estimated from the
  mean stellar age, the mean metallicity, and the star formation
  history of the galaxy.
\end{list}

Unfortunately, not all of these components can be determined
observationally.  For the SDSS data used in this analysis, only
$M_{dyn}$, $M_{\star,IMF}$, and $L$ are measurable quantities.  In
practice, $M_{\star}$ can only be robustly determined for resolved
stellar systems.  $M_{\star}$ values quoted for unresolved stellar
systems {\it always} rely upon stellar population modelling plus the
assumption of a single, constant IMF.  They are therefore really
values of $M_{\star,IMF}$, not a true measurement of $M_{\star}$.

Direct measurements of $M_{tot}$ are only possible where an
independent mass estimated can be derived.  Recent examples include
resolved kinematic observations of nearby early type galaxies
\citep{cappellari06} and strong galaxy-galaxy lenses
\citep{bolton07,bolton08,koopmans09} which show that the simple
dynamical estimator does an excellent job of reproducing the galaxy
masses derived from kinematics or lensing.  These studies suggest that
$M_{dyn} = M_{tot}$, with scatter of 0.06--0.07 dex standard deviation
around the one-to-one relation (cf. \citealt{cappellari06},
\citealt{bolton07}).  This level of variation in $M_{dyn}/M_{tot}$ is
clearly inadequate to explain the variation in $M_{dyn}/L$ in Figure
\ref{vdisp_mdyn}.  These observational results are further supported
by numerical simulations which show that galaxies with orbital
anisotropies large enough to move them off the FP are dynamically
unstable \citep{ciotti96, ciotti97, nipoti02}.

Assuming that $M_{dyn} = M_{tot}$ and bearing in mind that $M_{\star}$
is not a measurable quantity in unresolved stellar systems, equation
\ref{ml_equation4} simplifies to
\begin{equation}\label{ml_equation}
\frac{M_{dyn}}{L} = \frac{M_{dyn}}{M_{\star,IMF}} \times 
\frac{M_{\star,IMF}}{L},
\end{equation}
where both $M_{dyn}/M_{\star,IMF}$ and $M_{\star,IMF}/L$ are
measurable quantities.  Variations in $M_{\star,IMF}/L$ are due to
known stellar population effects, while variations in the quantity
$M_{dyn}/M_{\star,IMF}$ can be produced by variations either in the
IMF or in the dark matter fraction inside $R_e$.  Hereafter, we will
refer to the $M_{dyn}/M_{\star,IMF}$ term as the ``dark matter/IMF''
contribution with the understanding that variations in this term can
be attributed to either effect, or to both in combination.

Many previous studies have indicated that the stellar population term
($M_{\star,IMF}/L$) contributes not more than 1/2 of the observerd FP
tilt (e.g., \citealt{pahre95, prugniel97, pahre98, scodeggio98,
  mobasher99, padmanabhan04, trujillo04, gallazzi05, jun08,
  la_barbera08, hyde09b}, but see also \citealt{allanson09} who find
that using a single burst star formation history produces enough
variation in $M_{\star,IMF}/L$ to explain the full FP tilt).  This
then requires variations in $M_{dyn}/M_{\star,IMF}$ due to variations
in either the IMF (as proposed by \citealt{renzini93} and
\citealt{chiosi98}) or in the dark matter fraction inside $R_e$ (e.g.,
\citealt{renzini93, pahre01, padmanabhan04, gallazzi05, hyde09b}) to
produce the observed tilt of the FP.

The case for dark matter variation has recently gotten a boost from
simulations of galaxy mergers that include gaseous dissipation.
Several authors have shown that the degree of dissipation during
mergers determines the dark matter fraction within $R_e$ in merger
remnants (e.g., \citealt{robertson06, dekel06}).  Under realistic
conditions in which mergers of more massive galaxies have decreasing
amounts of gas, such simulations can reproduce the observed tilt of
the FP \citep{kobayashi05, robertson06, dekel06, covington08_thesis,
  hopkins08_fp}.  A limited quantity of further dissipationless
merging should preserve the tilted FP \citep{capelato95, dantas03,
  nipoti03, boylan-kolchin05, robertson06}.

It seems therefore that, while stellar population effects may
contribute to the tilt of the FP, further variation in the dark
matter/IMF term are required.  This paper shows that these two
contibutions to the FP tilt rotate the plane around different axes in
the 3D space, such that $M_{\star,IMF} \propto \sigma^{0.30}$, while
$M_{dyn}/M_{\star,IMF} \propto M_{dyn}^{0.24}$.  We also show that
dark matter/IMF variations are required to produce the {\it thickness}
of the FP: stellar population effects contribute some, but the
dominant component of variation must come from differences in the dark
matter/IMF term.

\section{Measuring $M_{\star,IMF}/L$}\label{mstar_l}

As discussed in section \ref{ml_contributions}, the quantities
$M_{dyn}$, $M_{\star,IMF}$, and $L$ are the only components of
equation \ref{ml_equation4} that can be determined for the SDSS data.
Of these, $M_{dyn}$ and $L$ are relatively direct measurements while
$M_{\star,IMF}$, the stellar mass estimate, comes from $L \times
M_{\star,IMF}/L$, where $M_{\star,IMF}/L$ is determined through
stellar population modelling.  In Paper II, we measured the mean age,
metallicity, and abundance patterns for the stacked spectra in 3D FP
space.  With these quantities determined, there are three remaining
factors that affect $M_{\star,IMF}/L$.  These are:
\begin{list}{}{}
\item[1.]{The stellar population model code used to compute
  $M_{\star,IMF}/L$.}
\item[2.]{The library of star formation histories used in the models.}
\item[3.]{Whether the stellar population modelling is done by fitting
  individual spectral absorption lines, or by fitting the broad-band
  galaxy spectral energy distributions (SEDs).}
\end{list}
This section addresses how each of these factors affect the
determination of $M_{\star,IMF}/L$, in order to motivate our choice of
models and to assess the extent to which our final conclusions depend
upon this choice.  Qualitatively, our conclusions turn out to be
independent of the exact choice of stellar population model and star
formation history used to determine $M_{\star,IMF}/L$, although
quantifying these relationships does depend on the choice of model.

\begin{figure*}[t]
\begin{center}
\includegraphics[width=0.9\linewidth]{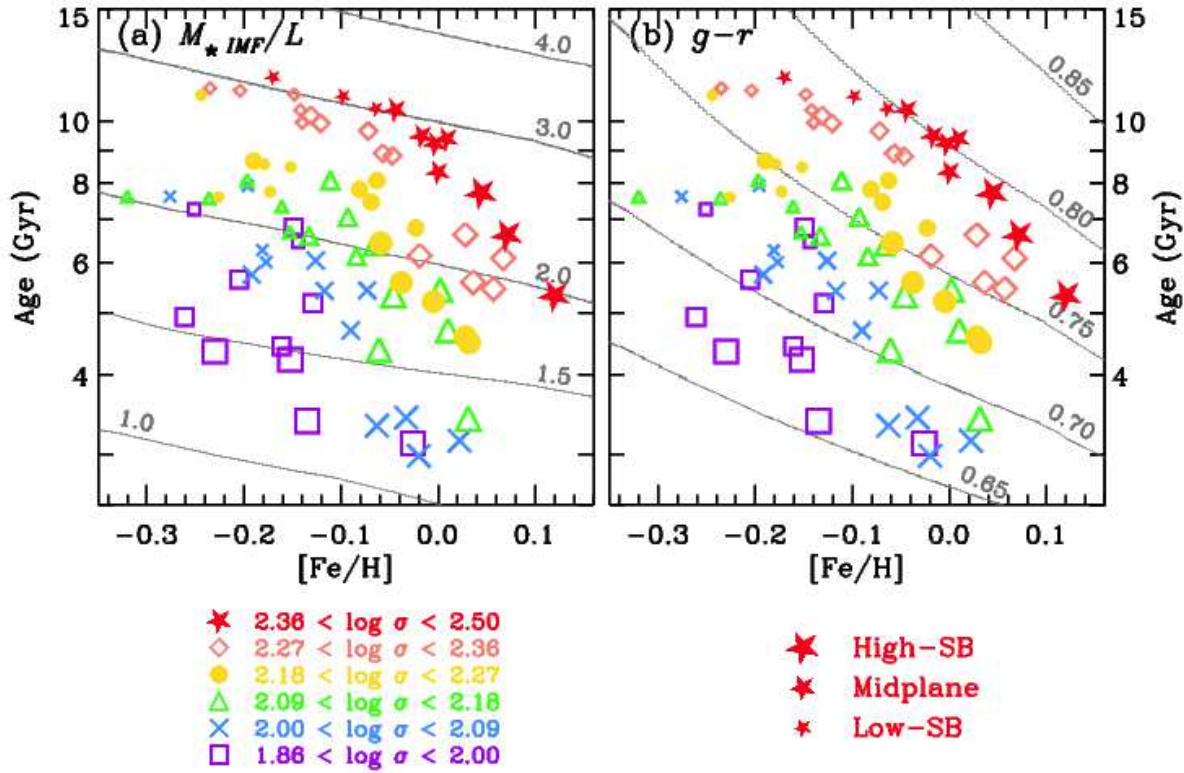}
\end{center}
\caption{Age and [Fe/H] values measured in the stacked spectra (3D FP
  binning).  Multiple symbols with the same size and color represent
  different values of $R_e$; only bins with small errors
  ($\Delta$[Fe/H] $< 0.05$ dex and $\Delta Age < 15$\%) are shown.
  Galaxies with higher values of $\sigma$ typically have older ages
  and higher [Fe/H] than galaxies with low values of $\sigma$.  At
  fixed $\sigma$, galaxies with high (low) surface brightness have
  younger (older) ages and higher (lower) [Fe/H] than galaxies at the
  same $\sigma$ that lie on the FP midplane (c.f., Paper II).  Lines
  of constant $M_{\star,IMF}/L$ (a) and lines of constant $g-r$ color
  (b) from BC03 models are overplotted as the gray lines.  From a
  given value of age and [Fe/H], the corresponding value of
  $M_{\star,IMF}/L$ and $g-r$ can be estimated.  $M_{\star,IMF}/L$
  depends strongly on age but only weakly on [Fe/H], while $g-r$
  depends strongly on both age and [Fe/H].  The anti-correlated
  variations in age and [Fe/H] at fixed $\sigma$ move galaxies nearly
  along lines of constant $g-r$ but spread the galaxies over a range
  of $M_{\star,IMF}/L$.}\label{hyperplane}
\end{figure*}

\citet{longhetti09} have recently addressed the first issue listed
above using a suite of stellar population models to predict
$M_{\star,IMF}/L$ based on optical and infra-red colors.  They examine
the dependence of $M_{\star,IMF}/L$ on age, metallicity, the IMF, and
the choice of stellar population model codes.  They explore several
popular model codes including those of \citet{bc03} and
\citet{maraston05}, as well as PEGASE \citep{fioc97}, and GRASIL
\citep{silva98}) and find that the different codes produce relatively
consistent $M_{\star,IMF}/L$ predictions.

If the ages and metallicities of the galaxies are known,
\citet{longhetti09} find that the largest variations in
$M_{\star,IMF}/L$ determinations are due to the choice of IMF, with a
zeropoint shift of $\sim 0.26$ dex between the \citet{salpeter55} and
\citet{chabrier03} IMFs and of $\sim 0.21$ dex between the Salpeter
and \citet{kroupa01} IMFs, in the sense that the Salpeter IMF predicts
larger values of $M_{\star,IMF}/L$ than either Chabrier or Kroupa.
Relative values of $M_{\star,IMF}/L$ between different galaxies are
robust to this effect, {\it as long as the IMF is the same for all
  galaxies.}

In the following sections, we examine the effects on $M_{\star,IMF}/L$
of various star formation histories, and compare SED-based estimates
of $M_{\star,IMF}/L$ with those derived from spectral absorption
features.  All of the models presented here are based on isochrones
computed for scaled-solar element abundance patterns.\footnote{Since
  the galaxies presented here typically have super-solar [Mg/Fe],
  these are not an ideal match to the data.  Mg-enhanced isochrones
  are shifted to cooler temperatures than scaled-solar isochrones, but
  a complicated interplay between different abundance effects makes it
  challenging to predict how various non-solar abundance patterns will
  affect $M_{\star,IMF}/L$ (e.g., \citealt{dotter07, lee10}).
  Understanding these effects in detail may well represent the next
  major step forward in stellar population modelling.}

\subsection{Single Burst Models for $M_{\star,IMF}/L$}\label{single_burst_ml}

As a reference point, we begin by computing our own values of
$M_{\star,IMF}/L$ using simple single-burst stellar population models
(SSPs).  Having mapped out the variations in mean age and metallicity
through 3D FP space in Paper II, we are now in a position to study how
these variations contribute to the observed variations in $M_{dyn}/L$.
The modelling process is illustrated in Figure \ref{hyperplane}, which
shows the values of mean, light-weighted age versus [Fe/H] determined
from the stacked galaxy spectra, as described in Paper II.  We have
overlaid lines of constant stellar population mass-to-light ratio
($M_{\star,IMF}/L$, panel a) and constant $g-r$ color (panel b)
derived from the single burst stellar population models of
\citet[hereafter BC03]{bc03} with a Chabrier IMF.  These curves make
it possible to read off model values of $M_{\star,IMF}/L$ and $g-r$
for each set of stacked spectra considered here.\footnote{The BC03
  models are computed for solar abundance ratios only, which is not an
  accurate reflection of the abundance pattern of our sample galaxies.
  Nevertheless, we show in Appendix \ref{compare_colors} that the BC03
  models, when chosen to match the SSP ages and [Fe/H] values from our
  stellar population modelling, are able to accurately reproduce the
  observed galaxy colors.}

The different colors and symbols encode the $\sigma$ values for each
stacked spectrum, while the symbol size encodes $\Delta I_e$, as
indicated.  The plot includes the various bins in $R_e$ but these are
not differentiated in the figure, as neither age nor [Fe/H] depend on
$R_e$ (see Paper II).  The figure illustrates the main conclusion of
Paper II: that the stellar populations of quiescent galaxies form a
two-parameter family.  The two dimensions of this family are such that
galaxies with higher values of $\sigma$ tend to have older ages and
higher [Fe/H] than galaxies with low $\sigma$, while at fixed $\sigma$
galaxies with lower surface brightnesses have older ages and lower
[Fe/H] than those with higher surface brightnesses.

It is interesting to note that, although galaxies at fixed $\sigma$
span a significant range in age and [Fe/H], this variation moves them
almost exactly along lines of constant $g-r$ color, as shown in Figure
\ref{hyperplane}b.  Thus, a process that attempts to use galaxy colors
to model stellar populations will not successfully distinguish between
the different populations at fixed $\sigma$.  This is due to a nearly
perfect alignment of the age-metallicity degeneracy in color with the
measured anti-correlated of age and [Fe/H] in galaxies at fixed
$\sigma$.  However, lines of constant $M_{\star,IMF}/L$ in Figure
\ref{hyperplane}a follow shallower trajectories compared to lines of
constant $g-r$ color, such that galaxies at fixed $\sigma$ span a
range of $M_{\star,IMF}/L$ values but not a range of colors.  Using
the full 5-band SDSS photometry does not give significantly improved
$M_{\star,IMF}/L$ values, as all optical colors follow similar
trajectories in age--[Fe/H] space.\footnote{There is a slight
  steepening of the lines of constant color as colors become bluer
  (e.g., lines of constant $u-g$ are steeper than $g-r$) but even the
  reddest $i-z$ color is not nearly as flat as the lines of constant
  $M_{\star,IMF}/L$ in Figure \ref{hyperplane}.}

The models shown here are based on single burst stellar population
models (SSPs), which are almost certainly not accurate descriptions of
the star formation histories of galaxies.  However, such simple models
are more transparent to interpret, and they provide a good reference
point for considering models with more complicated star formation
histories.  Appendix \ref{compare_colors} shows that the $g-r$ colors
derived from Figure \ref{hyperplane}b do an excellent job of
reproducing the observed galaxy colors, which gives us confidence that
the derived values of $M_{\star,IMF}/L$ are likewise reasonable.

\begin{figure*}[t]
\begin{center}
\includegraphics[width=0.9\linewidth]{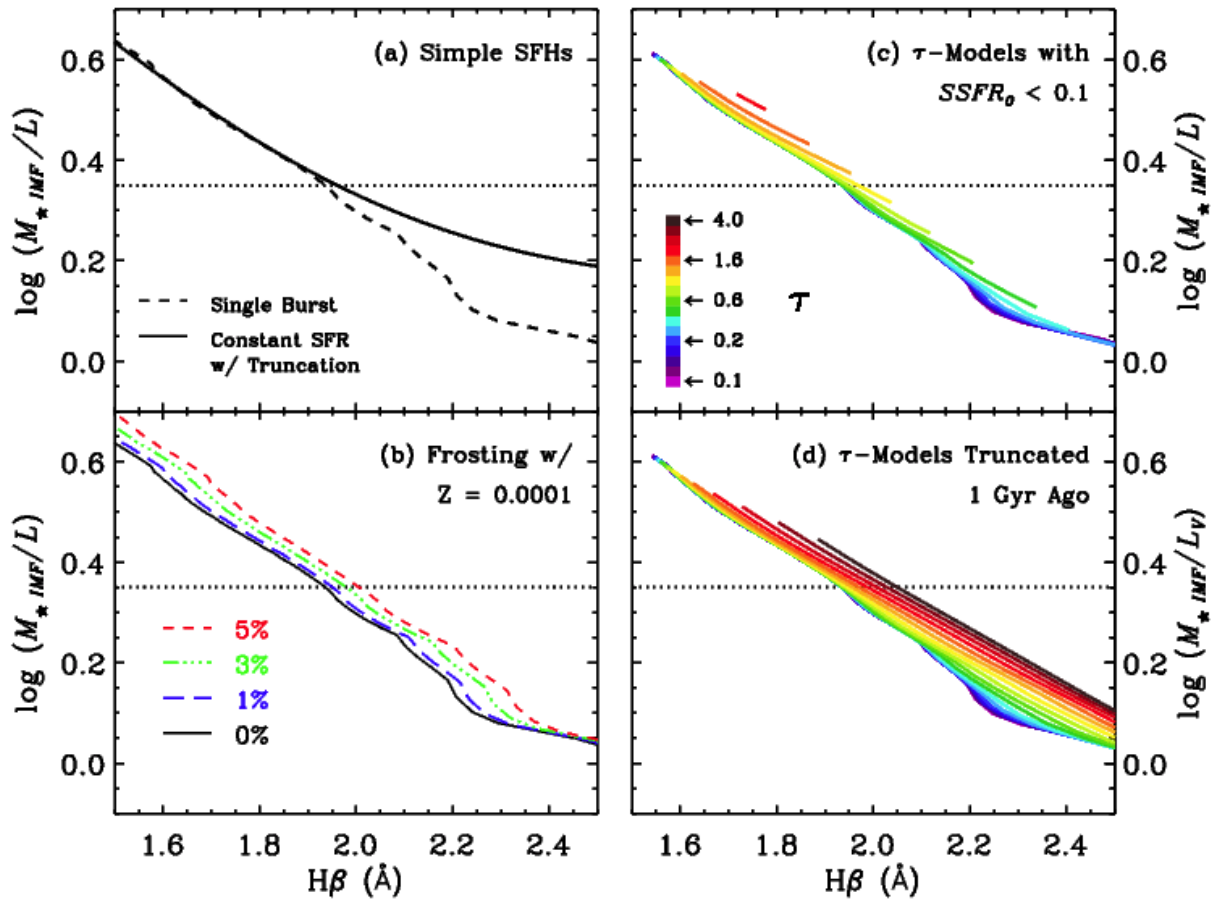}
\end{center}
\caption{A comparison of $M_{\star,IMF}/L$ values from models with
  different star formation histories.  All models are BC03 models with
  a Chabrier IMF and solar abundances. (a) Simple star formation
  histories: the dashed line shows $M_{\star,IMF}/L$ and H$\beta$
  values from single burst star formation histories, while the solid
  line shows values from models with constant star formation that
  begins at $z=10$ and truncates abruptly at various times.
  Differences in $M_{\star,IMF}/L$ due to star formation history
  become relevant for $\log (M_{\star,IMF}/L) < 0.35$, which
  corresponds to $t_{quench} = 3$ Gyr ago.  (b) Models with a frosting
  of metal-poor ($Z = 0.0001$) stars, with fractions as indicated.
  Small, very metal-poor populations produce slightly higher
  $M_{\star,IMF}/L$ for a given value of H$\beta$.  (c) $\tau$-models
  with $0.1 < \tau < 4.0$ Gyr and a range of initial formation
  redshifts ($0.25 < z_{f} < 40$).  Only models with SSFRs consistent
  with our sample galaxies are shown ($< 0.1 M_{\odot}$ yr$^{-1} /
  10^{11} M_{\odot}$).  These all have mass-to-light ratios consistent
  with the single burst models.  (d) $\tau$-models that have been
  abruptly truncated 1 Gyr before the time of observation.  This
  family of models produces a limited range of H$\beta$ values for a
  given value of $M_{\star,IMF}/L$, particularly for $\log
  (M_{\star,IMF}/L) < 0.35$.  }\label{hbeta_ml}
\end{figure*}

\subsection{Models for $M_{\star,IMF}/L$ with Complex
  Star Formation Histories}\label{tau_ml}

For galaxies with old stellar populations, the differences between
single burst values for $M_{\star,IMF}/L$ and those from extended star
formation histories are small.  However, galaxies with younger mean
light-weighted ages may harbour a bright young sub-population
obscuring a large population of older stars that contribute
substantially to $M_{\star}$ but very little to the integrated galaxy
light.  For these galaxies, the single burst estimates for younger
galaxies are almost certainly too low.

This is illustrated in Figure \ref{hbeta_ml}a, which plots
$M_{\star,IMF}/L$ against the H$\beta$ absorption line
strength.\footnote{ H$\beta$ is the major age discriminant used in the
  stellar population modeling of Paper II.}  The figure compares
single burst models (dashed line) to models with continuous star
formation (solid line) that begins at $z=10$, proceeds at a constant
star formation rate, then is abruptly truncated at various
times.\footnote{In order to isolate the effect of age distribution in
  the stellar populations, the constant star formation rate models, as
  well as the more complicated $\tau$-models considered in this
  section, are computed at Solar metallicity.}  These star formation
histories are indistinguishable for values of $\log (M_{\star,IMF}/L)
> 0.35$, corresponding to single burst ages $> 7$ Gyr or constant star
formation rates with truncation times $t_{t} > 3$ Gyr ago.  At lower
values of $M_{\star,IMF}/L$, single burst models can substantially
underestimate $M_{\star,IMF}/L$ if the true star formation history is
more extended.

In Figure \ref{hbeta_ml}c, we show H$\beta$ versus $M_{\star,IMF}/L$
for a set of exponentially declining star formation rates with
e-folding time $\tau$ (so-called ``$\tau$ models'').  For each $\tau$,
we compute a set of models in which the onset of star formation occurs
at different times, spanning a range formation redshifts with $0.25 <
z_f < 40$.  However, the majority of the models with $\tau > 0.5$
imply $z=0$ star formation rates that are too large for our quiescent
galaxy sample.  These are excluded from the set of viable models, as
explained below.

The sample presented here includes only galaxies with H$\alpha$
emission-line fluxes below a $2\sigma$ detection threshold.  We take
the $2\sigma$ error value as an upper limit on the intrinsic H$\alpha$
flux for each galaxy, then convert this limit into a star formation
rate following \citet[equation 2]{kennicutt98}.  Using the
\citet{gallazzi05} measurements of $M_{\star,IMF}$ for each galaxy
(explored in greater detail in \S\ref{spectral_vs_sed_ml}), we compute
an upper limit to the specific star formation rate (SSFR) for each
sample galaxy.  The vast majority (96\%) of our sample galaxies have
upper limits on their SSFRs that are below 0.1 $M_{\odot}$ yr$^{-1} /
10^{11} M_{\odot}$.  We therefore use this value to restrict the set
of $\tau$ models.

In Figure \ref{hbeta_ml}c, we show only models that have SSFRs at
$z=0$ that are less than 0.1 $M_{\odot}$ yr$^{-1} / 10^{11}
M_{\odot}$.  Low values of $\tau$ lead to models with short e-folding
times, which closely resemble single burst star formation histories.
At high values of $\tau$, only models in which star formation begins
at very early times have low enough SSFRs today to be included in the
sample.  The net effect is that the $M_{\star,IMF}/L$ values for a
given observed H$\beta$ line strength are nearly identical to the
single burst value, regardless of the values of $\tau$ or $z_f$.  For
$\tau$-model star formation histories, single burst models in fact
give a good approximation of the true value of $M_{\star,IMF}/L$.

However, models which include an abrupt truncation of star formation
can have substantially different values of $M_{\star,IMF}/L$ for a
given H$\beta$ line strength.  In Figure \ref{hbeta_ml}d, we modify
our library of $\tau$ models to include models that follow
exponentially declining star formation histories that are suddenly
truncated 1 Gyr before the time of observation.  These models show
some variation in H$\beta$ line strengths for a fixed value of
$M_{\star,IMF}/L$, depending on the star formation history.  The
spread is modest for galaxies with $M_{\star,IMF}/L > 0.35$ but is
larger for galaxies with lower $M_{\star,IMF}/L$.

Based on this suite of results, we infer that $M_{\star,IMF}/L$ values
above 0.35 are relatively insensitive to the details of the galaxy
star formation histories and are therefore reliable.  In contrast,
values of $M_{\star,IMF}/L$ below 0.35 may be off by 0.1--0.15 dex and
are more likely to be skewed too low for the youngest galaxies with
the lowest $M_{\star,IMF}/L$.\footnote{It should be noted that these
  statements apply to {\it smoothly varying} star formation histories.
  Non-continuous (i.e., bursty) star formation histories can also
  produce substantially different values of $M_{\star,IMF}/L$.
  However, in this analysis, we are examining stacks of many dozens of
  spectra, which means that the assumption of continuous star
  formation histories is reasonable.  This is one benefit of
  stacking.}

Finally, \citet{maraston09} have suggested that early type galaxies
may contain a small sub-population of very metal-poor stars.  They
find that models for single-metallicity old stellar populations cannot
match the color evolution of luminous red galaxies in the SDSS from
$z=0.7$ to $z=0.1$.  They demonstrate that adding a frosting (3\% by
mass) population of low metallicity ($Z = 0.0001$, i.e., [Fe/H]
$=-2.2$) stars makes it possible to match the observed color
evolution.  In Figure \ref{hbeta_ml}b, we illustrate the effect of
such a population on $M_{\star,IMF}/L$ and H$\beta$.  The 3\% frosting
population suggested by \citet{maraston09} results in
$M_{\star,IMF}/L$ values that are $< 0.05$ dex higher for a given
observed values of H$\beta$, which is comparable to the variation
produced by the $\tau$-model star formation histories considered here.

This leads to a key point.  {\it In all cases, the single burst models
  give a lower limit on $M_{\star,IMF}/L$}.  Single burst models in
fact give us a powerful tool because we can understand the biases in
measurements based upon them.  This gives us a benchmark against which
to compare other, more complicated determinations of
$M_{\star,IMF}/L$ for which the biases are less clear.

\begin{figure*}[t]
\begin{center}
\includegraphics[width=0.9\linewidth]{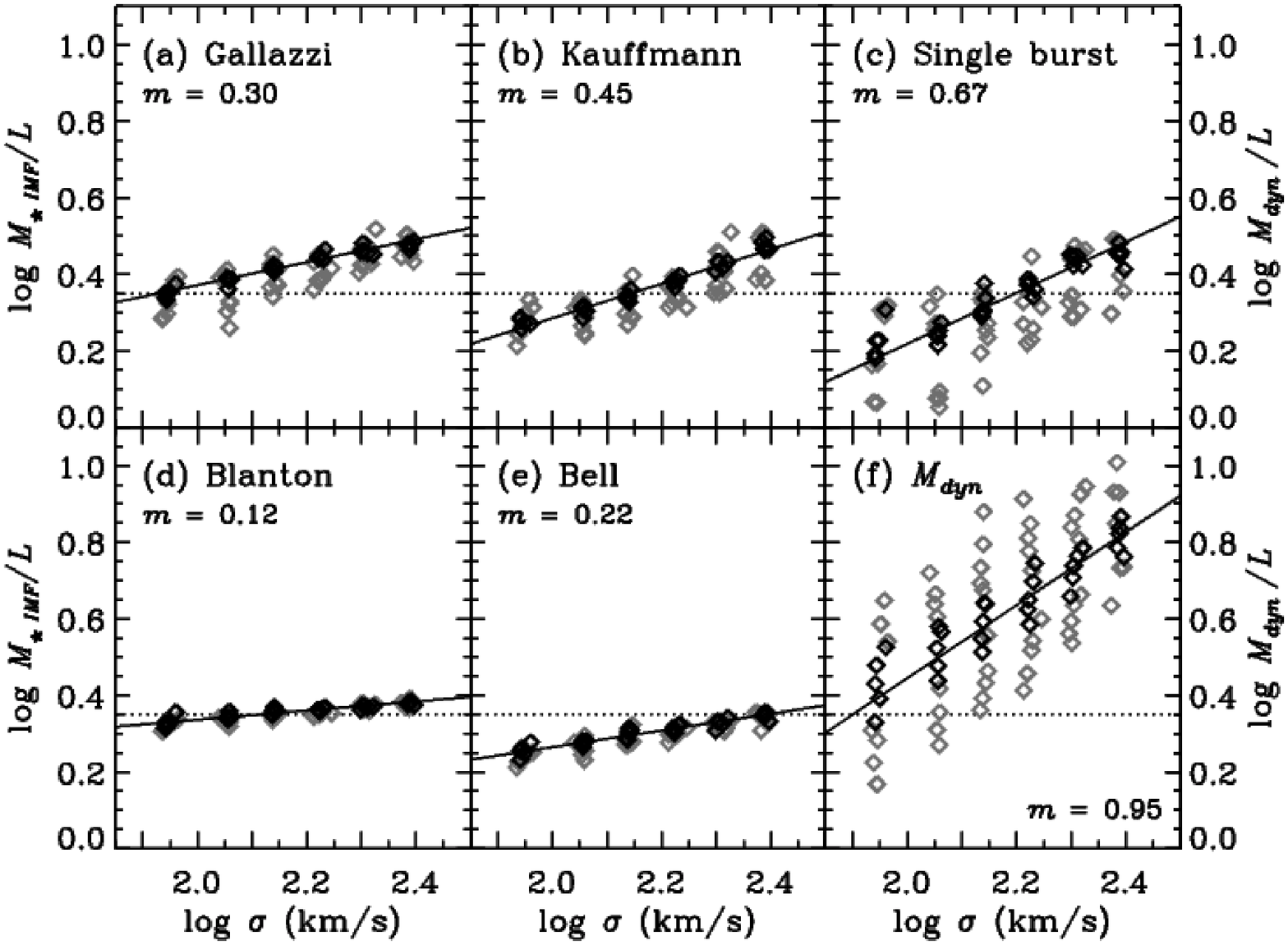}
\end{center}
\caption{(a-e) Various estimates of $M_{\star,IMF}/L$ for the galaxies
  in our sample as a function of $\sigma$.  Galaxy bins on the FP
  midplane are shown in black, while those on the low- and high-SB
  slices are shown in gray.  Panels (a-c) show values derived from
  spectral absorption features, including measurements from (a)
  \citet{gallazzi05}, (b) \citet{kauffmann03_mstar}, and (c) our
  single burst values.  Panels (d-e) show values based on fits to
  multi-band photometry, following the prescriptions of (d)
  \citet{blanton07} and (e) \citet{bell03}.  In each panel, solid
  lines show least squares fits of $M_{\star,IMF}/L$ onto $\sigma$ for
  the FP midplane galaxies only, weighted by the number of galaxies in
  each bin.  The slopes are indicated in each panel.  The dotted line
  at $M_{\star,IMF}/L = 0.35$ marks the point below which differences
  in the star formation history significantly affect the determination
  of $M_{\star,IMF}/L$ (see section \ref{tau_ml}).  (f) The values of
  $M_{dyn}/L$ from Figure \ref{vdisp_mdyn} are reproduced for
  comparison.  Regardless of the method used to measure
  $M_{\star,IMF}/L$, it clearly cannot provide either the full slope
  of $M_{dyn}/L$ with $\sigma$ (the ``$\sigma$ tilt'' of the FP) nor
  the spread in $M_{dyn}/L$ at fixed $\sigma$ (the ``thickness'' of
  the FP).  }\label{vdisp_ml}
\end{figure*}

\subsection{$M_{\star,IMF}/L$ from Spectral Absorption Lines
vs. Optical Colors}\label{spectral_vs_sed_ml}

Figure \ref{vdisp_ml} compares $M_{\star,IMF}/L$ from four different
sets of authors, two of which are computed based on fits to spectral
absorption features (\citealt{gallazzi05} and
\citealt{kauffmann03_mstar}) and two of which are based on optical
colors or multi-band photometry (\citealt{blanton07} and
\citealt{bell03}), to our benchmark single burst models.  All four of
these $M_{\star,IMF}/L$ measurements are available for the individual
SDSS galaxies used in our sample.  Each papers' version of
$M_{\star,IMF}/L$ uses slightly different stellar population models
and forms for the IMF.  We have corrected the zeropoints of the
\citet{kauffmann03_mstar} and \citet{bell03} values to match the
Chabrier IMF assumed by the \citet{gallazzi05}, \citet{blanton07}, and
our single burst models.

The \citet{kauffmann03_mstar} and \citet{gallazzi05} stellar mass
estimates for the individual SDSS galaxies were downloaded from the
derived data catalogs from SDSS studies at
MPA/JHU\footnote{http://www.mpa-garching.mpg.de/SDSS/}.
\citet{kauffmann03_mstar} determine $M_{\star,IMF}/L$ by fitting the
D$_n$4000 \citep{balogh99} and H$\delta_A$ \citep{worthey97} spectral
features to a library of models.  Their model star formation histories
comprise a set of exponentially declining star formation rates with
varying $\tau$, a range of total metallicities (always with the solar
abundance pattern), and superposed random bursts of star formation at
late times.  We have corrected the Kauffmann et al. stellar mass
measurements from SDSS Petrosian magnitudes to de Vaucouleurs
magnitudes to match the other mass estimates presented here.  The
\citet{gallazzi05} stellar masses (panel a) are modeled in a
comparable way using a similar library of star formation histories.
Gallazzi et al.\ use a larger number of spectral absorption features
in their fits; in particular, they include a number of redder features
at $\sim 5000$ {\AA}.

Figure \ref{vdisp_ml} shows that the two spectroscopic stellar mass
estimates agree very well for high-$\sigma$ galaxies, with
\citet{kauffmann03_mstar} finding somewhat lower $M_{\star,IMF}/L$ for
low-$\sigma$ galaxies.  This may be because their measurements only
use absorption lines around 4000~{\AA} and are therefore more strongly
influenced by young subpopulations of stars than are the Gallazzi et
al.\ estimates, resulting in underestimates of $M_{\star,IMF}/L$.
This difference produces a stronger slope to the $\log
(M_{\star,IMF}/L)$--$\log \sigma$ relation (gray line) in Kauffmann et
al.\ than in Gallazzi et al.

These values are compared to our single burst values in panel c, which
are similar to the Kauffmann et~al. values for most galaxy bins.  The
largest discrepancies are for galaxies with low $\sigma$ and low
$M_{\star,IMF}/L$, where the single burst estimates are significantly
lower than Kauffmann et~al.  These are precisely the galaxies for
which we expect the single burst approximation to break down,
resulting in single burst $M_{\star,IMF}/L$ values that are too low
(section \ref{tau_ml}).  

The photometry-based determinations of $M_{\star,IMF}/L$ look
different from all the spectroscopic values, as shown in Figures
\ref{vdisp_ml}d and \ref{vdisp_ml}e.  The stellar masses shown in
Figure \ref{vdisp_ml}d are computed using M.\ Blanton's IDL code {\it
  kcorrect} v.\ 4.1.4 \citep{blanton07}.  This code matches the SDSS
{\it ugriz} photometry (K-corrected to $z = 0$) to a linear
combination of a set of basis templates that have been determined to
span the space of SDSS galaxies.  The stellar masses in Figure
\ref{vdisp_ml}e are computed using the $B-V$ galaxy colors (determined
using {\it kcorrect}) and the conversions between galaxy color and
$M_{\star,IMF}/L$ given in Table 7 of \citet{bell03}.  The Bell et
al.\ conversions are based on fits of the full {\it ugrizK} combined
SDSS and 2MASS photometry for SDSS Early Data Release
\citep{stoughton02} galaxies to a set of models spanning a range of
star formation histories.  In both cases, the colors are computed from
photometry that has been matched to the SDSS $3''$ spectral fiber
aperture in order to provide consistency with the
spetroscopically-derived values of $M_{\star,IMF}/L$ (see appendix
\ref{compare_colors}).  These photometrically-derived values of
$M_{\star,IMF}/L$ show much weaker variation with $\sigma$ and much
smaller spread at fixed $\sigma$ than the spectroscopically-derived
values.

The increased spread in the spectroscopic values can be understood
from Figure \ref{hyperplane}b.  As discussed in section
\ref{single_burst_ml}, the lines of constant $g-r$ in the models of
Figure \ref{hyperplane}b run very nearly parallel to the lines of
constant $\sigma$ in the data.  The spectroscopic data are thus
capable of differentiating between different star formation histories
and different $M_{\star,IMF}/L$ values at fixed $\sigma$, which the
photometry-based values cannot do.

Throughout the rest of this analysis, we will use the
\citet{gallazzi05} values for $M_{\star,IMF}/L$ because they capture
the variation in $M_{\star,IMF}/L$ at fixed $\sigma$ (unlike the
photometry-based measurements) but are likely to be less biased toward
younger sub-populations than the \citet{kauffmann03_mstar} and
especially the single burst values.

\subsection{$M_{dyn}/L$ versus $M_{\star,IMF}/L$}\label{mdyn_vs_mstar} 

As a final point of comparison, Figure \ref{vdisp_ml}f shows
$M_{dyn}/L$ as a function of $\sigma$.  It is clear that none of the
$M_{\star,IMF}/L$ measurements (panels a--e) resemble $M_{dyn}/L$.
{\it No matter which method is used to measure $M_{\star,IMF}/L$,
  stellar population effects cannot reproduce the observed variation
  in $M_{dyn}/L$}.

In the first place, none of the $M_{\star,IMF}/L$ measurements have a
slope with $\sigma$ that is as steep as the slope of the observed
$M_{dyn}/L$--$\sigma$ relation.  This means that stellar population
effects are not enough to explain the tilt of the FP, as has been
demonstrated by numerous previous authors (e.g., \citealt{pahre95,
  prugniel96, pahre98, scodeggio98, mobasher99, padmanabhan04,
  trujillo04, jun08, la_barbera08, hyde09b}).\footnote{In contrast to
  our results and other previous studies, \citet{allanson09} find that
  the assumption of single burst star formation histories predicts a
  strong enough scaling of $M_{\star,IMF}/L$ with $\sigma$ to account
  for the entire $\sigma$ tilt of the FP.  While our results agree
  that single burst models produce substantially stronger slopes than
  those computed with more complex star formation histories, we find
  that even single burst models cannot produce the entire $\sigma$
  tilt of the plane. A comparison between our Figure \ref{vdisp_ml}
  and their Figure 12 suggests that this difference is due to
  differences in the slope of the observed $M_{dyn}/L$-$\sigma$
  relation (they find $M_{dyn}/L \propto \sigma^{0.79}$ while we find
  $M_{dyn}/L \propto \sigma^{0.95\pm0.04}$).  Our single burst stellar
  population modelling results agree with theirs; they find
  $M_{\star,IMF}/L \propto \sigma^{0.66}$ for a single burst model,
  while we find $M_{\star,IMF}/L \propto \sigma^{0.67\pm0.04}$.}

In the second place, it is clear from Figure \ref{vdisp_ml} that
$M_{\star,IMF}/L$ variations cannot provide enough spread at fixed
$\sigma$ to match the observed variation in $M_{dyn}/L$.  {\it This
  new result indicates that stellar populations contribute little to
  the thickness of the FP}, despite the fact that residuals from the
FP appear to be correlated with stellar population properties
(\citealt{forbes98, wuyts04, gargiulo09}, Paper II).  This is because
the stellar population differences through the thickness of the FP
show an {\it anti}-correlation between age and [Fe/H].  This
anti-correlation, while not exactly parallel to lines of constant
$M_{\star,IMF}/L$ (see Figure \ref{hyperplane}), is in a similar
direction and therefore limits the differences in $M_{\star,IMF}/L$
that result from the stellar population differences at fixed $\sigma$.

The marked difference of Figure \ref{vdisp_ml}f from Figure
\ref{vdisp_ml}a--e {\it requires} that there be variations in
$M_{dyn}/M_{\star}$ (the dark matter fraction) and/or in
$M_{\star}/M_{\star,IMF}$ (the IMF) among quiescent galaxies.  At
least one of these quantities must vary {\it along} the FP to
reproduce the observed $\sigma$ tilt of the plane.  One or both of
them must also vary {\it through the thickness} of the FP.
Furthermore, these variations likely dominate over stellar population
variations both along and through the FP.

\section{Mass-to-Light Ratios in 3D FP Space}\label{ml_in_3d}

Having quantified the variation of $M_{\star,IMF}/L$ for our sample
galaxies, as well as the systematic uncertainties in $M_{\star,IMF}/L$
measurements, we can now examine how the stellar population term
contributes to the total observed variation of $M_{dyn}/L$ throughout
3D FP space.  In essence, we will be {\it dividing out} the variations
due to stellar populations and thereby mapping how the remaining term
($M_{dyn}/M_{\star,IMF}$) varies over the plane and through it.

Section \ref{fp_tilt} focuses on the FP midplane, examining how
$M_{\star,IMF}/L$ and $M_{dyn}/M_{\star,IMF}$ contribute separately to
the tilt of the FP.  We show that, not only does the tilt of the FP
require contributions from both $M_{\star,IMF}/L$ and
$M_{dyn}/M_{\star,IMF}$, but that these two components of the tilt
rotate the plane {\it around different axes} in the 3D parameter
space.  Section \ref{fp_thickness} then maps both these quantities in
a cross-section through the FP, examining how each contributes to the
thickness of the FP.  We show that the thickness of the FP is
dominated by variations in $M_{dyn}/M_{\star,IMF}$, with
$M_{\star,IMF}/L$ contributing only $\sim 22$\% of the variation.
Finally, section \ref{fp_fits} distills the quantitative results of
this analysis by presenting fitting functions for the total
mass-to-light variation ($M_{dyn}/L_V$), the stellar population
contribution ($M_{\star,IMF}/L$), the dark matter/IMF contribution
($M_{dyn}/M_{\star,IMF}$), and the associated surface mass density of
stars ($\Sigma_{\star,IMF}$) as functions of the three FP variables:
$\sigma$, $R_e$, and $\Delta I_e$.

\begin{figure*}[t]
\includegraphics[width=1.0\linewidth]{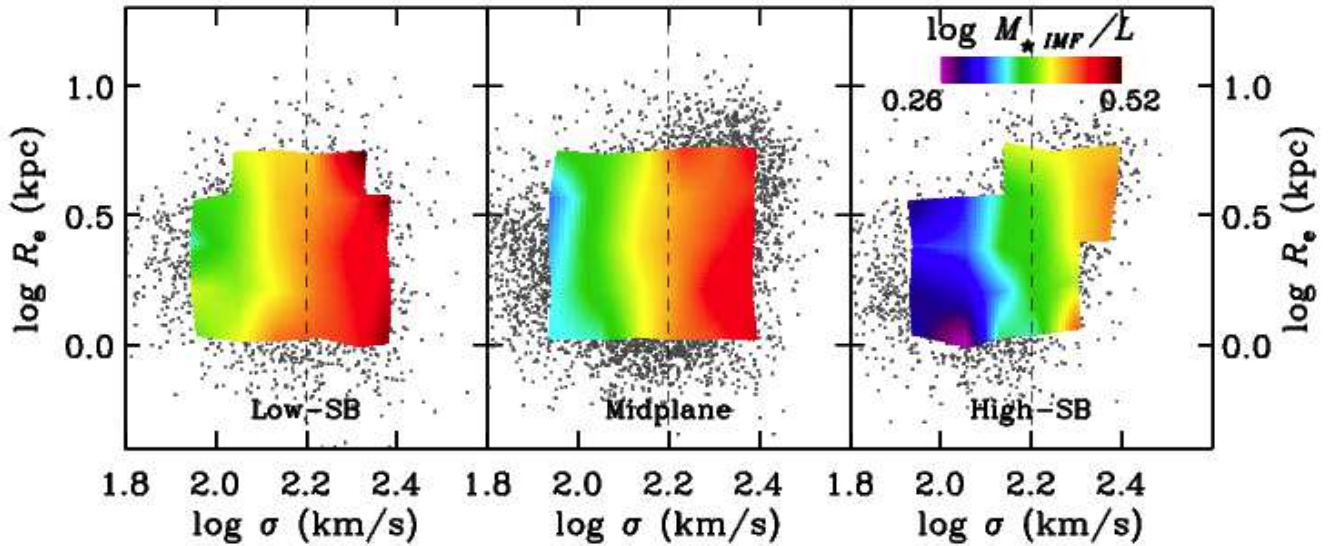}
\caption{The FP tilt contribution from stellar population effects,
  mapped throughout 3D FP space.  Data points in each panel represent
  the sample galaxies divided into three slices in surface brightness,
  as in Figure \ref{fp_bins}.  The overlaid color contours show
  variations in $M_{\star,IMF}/L$ in this space, measured using the
  values of \citet{gallazzi05}.  On the FP midplane (center panel),
  lines of constant $M_{\star,IMF}/L$ follow lines of constant
  $\sigma$ (dashed line); $M_{\star,IMF}/L$ is independent of $R_e$.
  This indicates that $M_{\star,IMF}/L$ {\it is a function of only
    $\sigma$} ($M_{\star,IMF}/L = f(\sigma)$) on the FP.  The low- and
  high-surface-brightness slices show the same effect, with lines of
  constant $M_{\star,IMF}/L$ running vertically.  However, they show
  zeropoint shifts from the midplane, with the low- and
  high-surface-brightness galaxies have systematically higher and
  lower $M_{\star,IMF}/L$, respectively.  These trends in
  $M_{\star,IMF}/L$ are as predicted from the stellar population maps
  in Paper II.  }\label{ml_maps}
\end{figure*}

\subsection{The Tilt of the FP in 3D}\label{fp_tilt}

Figure \ref{ml_maps} maps how the stellar population $M_{\star,IMF}/L$
component varies over the FP using the bins presented in Figure
\ref{fp_bins}.  The three panels show the three slices in surface
brightness, including the midplane and the slices above and below the
plane.  The overlaid contours indicate the median $M_{\star,IMF}/L$
values for each bin in FP space, using the \citet{gallazzi05}
measurements.  The contours are constructed by plotting the median
value of $M_{\star,IMF}/L$ for each bin at the point corresponding to
the median values of $\sigma$ and $R_e$ for the bin.  The values of
$M_{\star,IMF}/L$ are then linearly interpolated between the $6 \times
5$ grid of bins in each surface brightness slice to produce a
continuous map of $M_{\star,IMF}/L$ along each slice.

Looking first at the FP midplane (center panel), it is clear that
$M_{\star,IMF}/L$ depends on $\sigma$ but is independent of $R_e$.
Lines of constant $M_{\star,IMF}/L$ follow lines of constant $\sigma$
(dashed line).  Therefore, we find that $M_{\star,IMF}/L$ is a
function only of $\sigma$ on the FP.  This is expected from Paper II,
where we showed that all stellar population properties (age, [Fe/H],
[Mg/H], and [Mg/Fe]) follow lines of constant $\sigma$ on the FP and
are essentially independent of $R_e$.  It therefore follows that the
known stellar population contribution to $M_{dyn}/L$ should also
follow lines of constant $\sigma$.  

The low- and high-surface-brightness panels above and below the
midplane show the same behavior: $M_{\star,IMF}/L$ again follows lines
of constant $\sigma$ and is independent of $R_e$.  However, the
zeropoint of the relation shifts such that galaxies in the
low-surface-brightness slice have {\it higher} $M_{\star,IMF}/L$ for a
given $\sigma$ than their counterparts on the FP midplane.  Similarly,
galaxies in the high-surface-brightness slice have {\it lower}
$M_{\star,IMF}/L$ for a given $\sigma$.  This demonstrates that there
are $M_{\star,IMF}/L$ variations through the {\it thickness} of the FP
as well, although they are relatively weak.  We will study these
variations through the thickness of the FP in more detail in
\S\ref{fp_thickness}.

\begin{figure*}[t]
\includegraphics[width=1.0\linewidth]{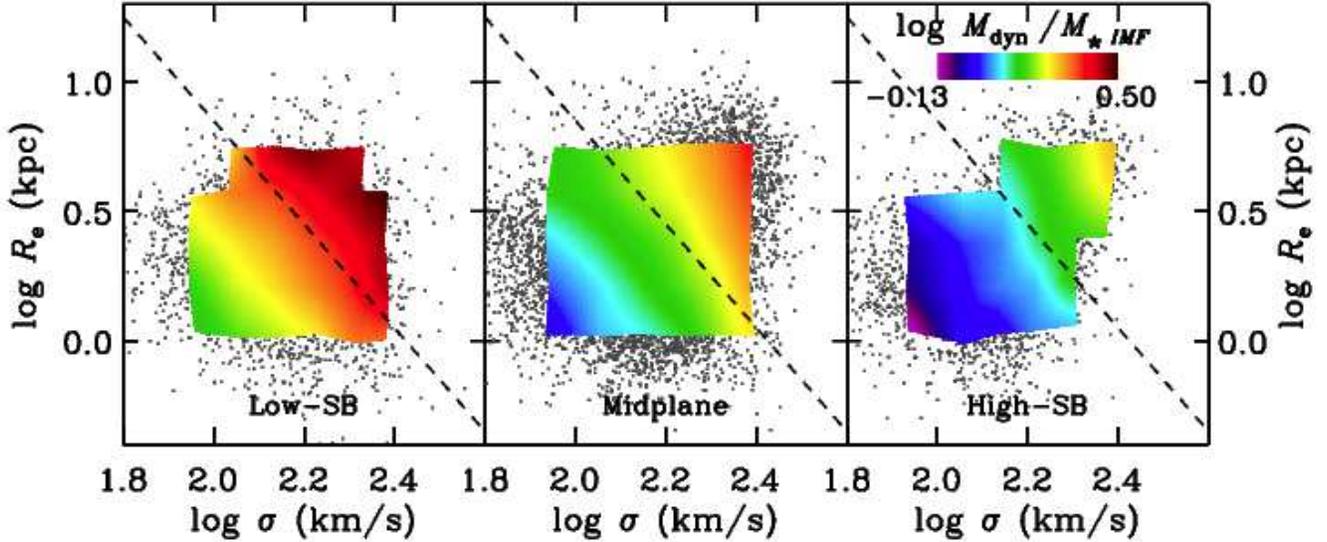}
\caption{The FP tilt contribution from varying dark matter/IMF
  effects, mapped throughout 3D FP space.  Data points in each panel
  represent the sample galaxies, divided into three slices in surface
  brightness, as in Figure \ref{fp_bins}.  The overlaid color contours
  show variations in $M_{dyn}/M_{\star,IMF}$ in this space.  On the FP
  midplane (center panel), lines of constant $M_{dyn}/M_{\star,IMF}$
  approximately follow lines of constant $M_{dyn}$ (dashed line) such
  that $M_{dyn}/M_{\star,IMF} = f(M_{dyn})$ on the FP.  The two
  components of the FP tilt therefore rotate the plane around
  different axes.  The low-surface-brightness
  (high-surface-brightness) slice shows the same effect, but with a
  zeropoint shift such that $M_{dyn}/M_{\star,IMF}$ values are higher
  (lower) than the midplane values.  }\label{mm_maps}
\end{figure*}

With $M_{\star,IMF}/L$ now mapped, we can divide it out to isolate the
map of $M_{dyn}/M_{\star,IMF}$ alone.  We do this by calculating the
ratio of the two known quantitites $M_{dyn}/L$ and $M_{\star,IMF}/L$.
The resultant map, which shows the dark matter/IMF contribution
$M_{dyn}/M_{\star,IMF}$, is shown in Figure
\ref{mm_maps}.\footnote{The values of $M_{dyn}/M_{\star,IMF}$ run from
  $-0.13$ to $+0.50$.  Negative values of $M_{dyn}/M_{\star,IMF}$ are
  clearly nonsensical if $M_{dyn} = 5 \sigma^2 R_e / G$ gives a true
  representation of the total mass and the Chabrier IMF is a good
  approximation to the true IMF in these galaxies; a galaxy cannot
  contain less total mass than its stellar mass.  However, modest
  changes to the assumed IMF (e.g., \citealt{longhetti09}) or to the
  constant used to compute $M_{dyn}$ will produce zeropoint shifts of
  the needed magnitude ($\sim0.1$ dex).  In this work, we have focused
  on trends through FP space, for which the zeropoint is not relevant,
  rather than on absolute measurements of $M_{\star,IMF}/L$.  It is
  worth noting however that, as computed here, the single burst values
  of $M_{\star,IMF}/L$ are the only ones for which all galaxy bins
  give $M_{\star,IMF} \leq M_{dyn}$.}  The data and contours are
constructed in the same way as Figure \ref{ml_maps}.

A major result of this paper is illustrated in Figure \ref{mm_maps}:
lines of constant $M_{dyn}/M_{\star,IMF}$ on the FP midplane (center
panel) follow a noticeably different slope than the lines of constant
$M_{\star,IMF}/L$ in Figure \ref{ml_maps}.  The new lines run
approximately parallel to lines of constant $M_{dyn}$ (dashed line,
with $M_{dyn} \propto \sigma^2 R_e$), indicating that on the FP
midplane $M_{dyn}/M_{\star,IMF}$ is a function of $M_{dyn}$.  Similar
behavior is seen in the low- and high-surface-brightness panels.  Both
\citet{gallazzi05} and \citet{hyde09b} showed a correlation between
$M_{\star,IMF}$ and $M_{dyn}$ that was strong, but this is the first
time that it has been shown to be the {\it best} correlation, by
mapping $M_{dyn}/M_{\star,IMF}$ explicitly in 3D FP space.  

By examing the two contributions to the FP tilt in 3D space, we have
revealed that {\it the two suspected causes of FP tilt rotate the
  plane around different axes.}  The stellar population contribution
has $M_{\star,IMF}/L \approx f(\sigma)$, while the dark matter/IMF
contribution has $M_{dyn}/M_{\star,IMF} \approx g(M_{dyn}) =
g(\sigma^2 R_e)$.  This indicates that the tilt of the FP cannot be
parameterized as a simple power-law function of $M_{dyn}$ or
$M_{\star}$ or $L$ or $\sigma$.  Instead, it takes the functional form
$f(\sigma) \times g(M_{dyn})$.

Figure \ref{mm_maps} also illustrates a second major result of this
paper.  Lines of constant $M_{dyn}/M_{\star,IMF}$ run parallel to
lines of constant $M_{dyn}$ in all three panels, but there is a
zeropoint shift between the slices.  The low-surface-brightness and
high-surface-brightness slices have respectively higher and lower
values of $M_{dyn}/M_{\star,IMF}$ for a given $\sigma$ than do their
counterparts on the FP midplane.  The sign is the same as the shift of
$M_{\star,IMF}/L$ seen in the three panels of Figure \ref{ml_maps}.
Both 3D maps reveal that the stellar population effects and the dark
matter fraction/IMF effects which contribute the the FP tilt
contribute also to the thickness of the FP as well.  Section
\ref{fp_thickness} explores this second trend in more detail.

\begin{figure*}[t]
\begin{center}
\includegraphics[width=0.9\linewidth]{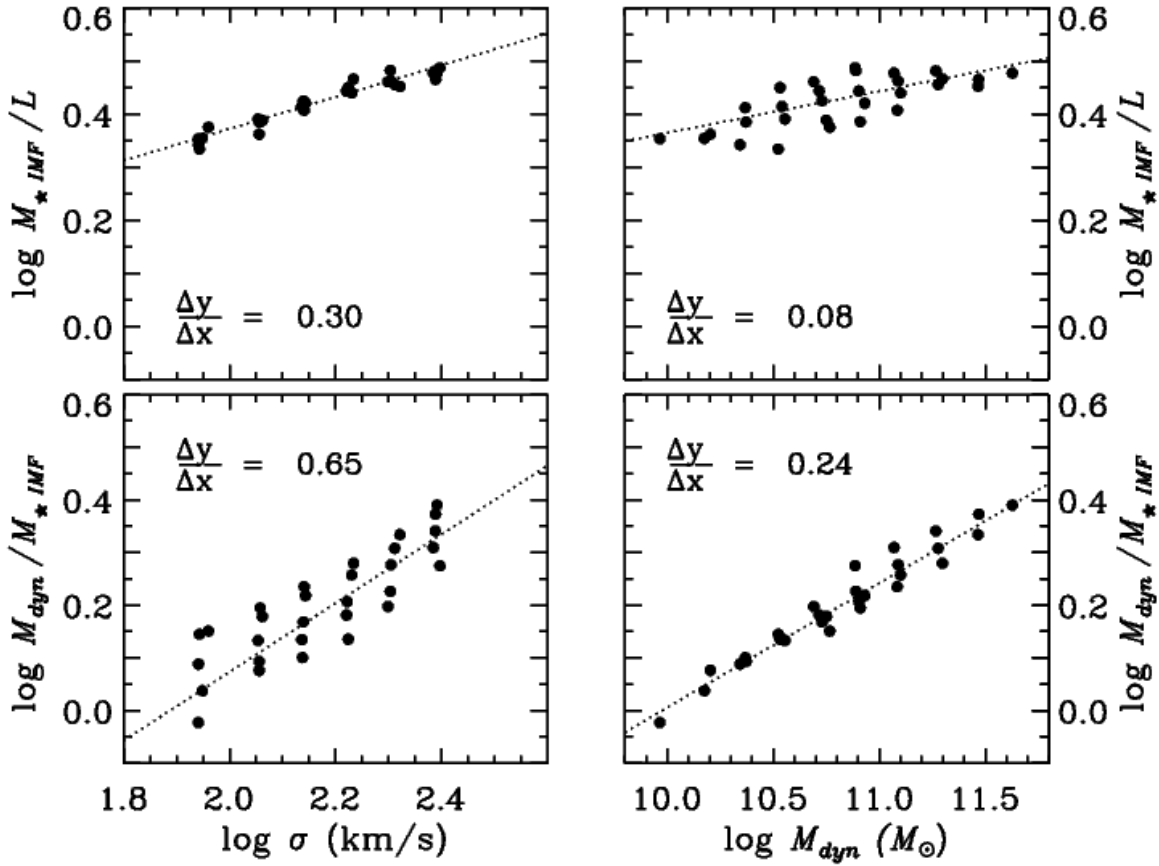}
\end{center}
\caption{The two tilts of the Fundamental Plane. Only galaxy bins from
  the FP midplane are shown. {\it Left:} The stellar population (upper
  panel) and dark matter/IMF (bottom panel) contributions to the
  $\sigma$ tilt of the FP.  The stellar population component is a
  tight function of $\sigma$, with $M_{\star,IMF}/L \propto
  \sigma^{0.30}$ and scatter of 0.01 dex.  The dark matter/IMF
  component shows more scatter when plotted against $\sigma$, with
  $M_{dyn}/M_{\star,IMF} \propto \sigma^{0.65}$ and scatter of 0.05
  dex.  Comparing the slopes of the two relations indicates that dark
  matter/IMF effects contribute 2/3 of the $\sigma$ tilt on the FP.
  {\it Right:} The stellar population (upper panel) and dark
  matter/IMF (lower panel) contributions to the total tilt of the FP.
  Here, the tightness of the relations is the opposite way round: the
  stellar population term shows 0.03 dex of scatter at fixed $M_{dyn}$
  with $M_{\star,IMF}/L \propto M_{dyn}^{0.08}$, while the dark
  matter/IMF term shows only 0.02 dex of scatter and has
  $M_{dyn}/M_{\star,IMF} \propto M_{dyn}^{0.24}$.  Dark matter/IMF
  effects contribute 3/4 of the total tilt of the FP.
}\label{two_tilts}
\end{figure*}

Focusing for the time being on the FP midplane alone, we quantify the
contributions of the two tilt components in Figure \ref{two_tilts}.
The left panels show the stellar population (upper left) and dark
matter/IMF (lower left) contributions to the $\sigma$ tilt of the FP.
Only the galaxy bins on the FP midplane slice are included.
$M_{\star,IMF}/L$ is a very tight function of $\sigma$.  To quantify
the stellar population contribution to the $\sigma$ tilt of the FP, we
calculate the ratio $\Delta \log (M_{\star,IMF}/L) / \Delta \log
\sigma$.  This is the first of several such ratios we will discuss;
hereafter we refer to them as ``$\Delta$--$\Delta$'' relations.  Note
that the numeric value of $\Delta \log (M_{\star,IMF}/L) / \Delta \log
\sigma$ is equivalent to the quantity $\mu$ in the relation
$M_{\star,IMF}/L \propto \sigma^{\mu}$.  A linear least-squares fit,
weighted by the number of galaxies in each bin, gives $M_{\star,IMF}/L
\propto \sigma^{0.30}$ and scatter of only 0.01 dex.  Meanwhile,
$M_{dyn}/M_{\star,IMF}$ is a less tight function of $\sigma$, with
$M_{dyn}/M_{\star,IMF} \propto \sigma^{0.65}$ and scatter of 0.05 dex.
Comparing the two slopes shows that {\it dark matter/IMF variations
  contribute twice as much to the $\sigma$ tilt of the FP as do
  stellar population variations.}

We can instead choose to parameterize the FP tilt as a function of
$M_{dyn}$ (i.e., the ``total tilt'' of the FP).  In this case, the
$\Delta$--$\Delta$ relation between $M_{dyn}/M_{\star,IMF}$ and
$M_{dyn}$ (lower right panel) is significantly tighter than that
between $M_{dyn}/M_{\star,IMF}$ and $\sigma$ (lower left panel), with
$M_{dyn}/M_{\star,IMF} \propto M_{dyn}^{0.24}$ and scatter of only
0.02 dex.  In contrast, $M_{\star,IMF}/L$ shows a less tight relation
with $M_{dyn}$ (upper right panel), giving $M_{\star,IMF}/L \propto
M_{dyn}^{0.08}$ and scatter of 0.03 dex.  Comparing the slopes of the
two relations shows that {\it dark matter/IMF variations contribute
  three times as much to the total tilt of the FP as do stellar
  population variations.}

Figures \ref{ml_maps}--\ref{two_tilts} use the \citet{gallazzi05}
values of $M_{\star,IMF}/L$, but any other choice gives qualitatively
similar results: in the FP midplane, lines of constant
$M_{\star,IMF}/L$ follow lines of constant $\sigma$, while lines of
constant $M_{dyn}/M_{\star,IMF}$ follow lines of constant $M_{dyn}$.
However, {\it quantifying} the various $\Delta$--$\Delta$ relations
depends on the choice of $M_{\star,IMF}/L$.  

These differences are summarized in Table \ref{quant_table}.  The top
section of the table shows the stellar population and dark matter/IMF
contributions to the $\sigma$ tilt of the FP, measured using the five
different estimates of $M_{\star,IMF}/L$ that were discussed in
\S\ref{spectral_vs_sed_ml}.  The \citet{gallazzi05} values are those
illustrated in the left panels of Figure \ref{two_tilts}; values using
other $M_{\star,IMF}/L$ estimates are computed in the same way.
Notice that, depending on the chosen method for estimating
$M_{\star,IMF}/L$, the stellar population contribution to the $\sigma$
tilt of the FP ranges from 13--70\%, with our preferred value falling
in the middle of the range at 32\%.

The middle section of Table \ref{quant_table} quantifies the stellar
population and dark matter/IMF contributions to the total tilt of the
FP, based on the five different estimates of $M_{\star,IMF}/L$.  For
any given estimate, stellar populations contribute less to the total
tilt of the FP than they do to the $\sigma$ tilt, ranging from
13--58\%, with our preferred value falling in the middle at 25\%.

\begin{deluxetable*}{lccccc}
\tabletypesize{\footnotesize}
\tablecaption{Stellar Population Contribution to the ``$\sigma$
  Tilt'', Total Tilt, and the Thickness of the FP for Various Methods
  of Measuring $M_{\star,IMF}$\label{quant_table}}
\tablehead{
\colhead{} &
\colhead{Gallazzi\tablenotemark{*}} &
\colhead{Kauffmann\tablenotemark{\dag}} &
\colhead{Single Burst\tablenotemark{\ddag}} &
\colhead{Blanton\tablenotemark{\S}} &
\colhead{Bell\tablenotemark{\P}}  
}
\startdata
\multicolumn{6}{c}{The $\sigma$ Tilt of the FP} \\
\hline \\
\vspace{0.15in} \large{$\frac{\Delta \log (M_{\star,IMF}/L)}{\Delta \log \sigma}$}
&0.30 &0.45 &0.67 &0.12 &0.22 \\
\vspace{0.15in} \large{$\frac{\Delta \log (M_{dyn}/M_{\star,IMF})}{\Delta \log
    \sigma}$} 
&0.65 &0.50 &0.29 &0.83 &0.73 \\
\vspace{0.15in} Fraction from Stellar Pops &32\% &47\% &70\% &13\% &23\% \\
\hline\hline \\
\multicolumn{6}{c}{The Total ($M_{dyn}$) Tilt of the FP} \\
\hline \\
\vspace{0.15in} \large{$\frac{\Delta \log (M_{\star,IMF}/L)}{\Delta \log M_{dyn}}$}
&0.08 &0.13 &0.18 &0.04 &0.07 \\
\vspace{0.15in} \large{$\frac{\Delta \log (M_{dyn}/M_{\star,IMF})}{\Delta \log
    M_{dyn}}$} 
&0.24 &0.19 &0.13 &0.28 &0.25 \\
\vspace{0.15in} Fraction from Stellar Pops &25\% &41\% &58\% &13\% &22\% \\
\hline\hline \\
\multicolumn{6}{c}{The Thickness of the FP} \\
\hline \\
\vspace{0.15in} \large{$\frac{\Delta \log (M_{\star,IMF}/L)}{\Delta \log I_e}$}
&$-0.22$ &$-0.20$ &$-0.54$ &$-0.02$ &$-0.02$ \\
\vspace{0.15in} \large{$\frac{\Delta \log (M_{dyn}/M_{\star,IMF})}{\Delta \log I_e}$} 
&$-0.79$ &$-0.81$ &$-0.48$ &$-1.00$ &$-0.99$ \\
\vspace{0.15in} Fraction from Stellar Pops &22\% &20\% &53\% &2\% &2\% \\
\enddata
\tablecomments{ \\
{*}{Values computed using the \citet{gallazzi05}
  measurements of $M_{\star,IMF}$.  These are our preferred values, as
discussed in section \ref{spectral_vs_sed_ml}.} \\
{\dag}{Values computed using the \citet{kauffmann03_mstar}
  measurements of $M_{\star,IMF}$.} \\
{\ddag}{Values computed using the single burst 
  measurements of $M_{\star,IMF}$ from section \ref{single_burst_ml}.} \\
{\S}{Values computed using the \citet{blanton07}
  measurements of $M_{\star,IMF}$.} \\
{\P}{Values computed using the \citet{bell03}
  measurements of $M_{\star,IMF}$.} }
\end{deluxetable*}

For both the $\sigma$ tilt and the total tilt, using the single burst
values gives the strongest variations in $M_{\star,IMF}/L$.  This is
due to the fact that single burst models likely underestimate
$M_{\star,IMF}/L$ for galaxies with the lowest values of
$M_{\star,IMF}/L$, making the trend look stronger than it actually is
(see section \ref{tau_ml}).  In contrast, the photometry-based
determinations of $M_{\star,IMF}/L$ from \citet{blanton07} and
\citet{bell03} produce very little variation in
$M_{\star,IMF}/L$---all galaxy bins have nearly identical values---so
that stellar population effects contribute very little to the $\sigma$
tilt or total tilt using these estimates.  

The key point is that, regardless of which method is used to estimate
$M_{\star,IMF}/L$, both stellar population and dark matter/IMF effects
are {\it required} to explain the tilt of the FP.  Furthermore, these
two components of the FP tilt rotate the plane around different axes.

Recently, \citet{treu10} have argued using entirely independent
methods that $M_{dyn}/M_{\star,IMF}$ (which they parameterize as an
``IMF mismatch parameter'' $\alpha$) varies as a function of $\sigma$,
such that higher-$\sigma$ galaxies have higher $\alpha$.  They find
$\log (M_{dyn}/M_{\star,IMF}) \propto (1.20 \pm 0.25) \log \sigma$ and
argue that these variations are either due to differences in the IMF
as a function of $\sigma$, or differences in the central profile of
the host dark matter halo.  The slope of the trend they find is
somewhat larger than what is presented here; their values of
$M_{\star,IMF}$ are inferred from multiband photometry and are thus
best compared to the Blanton et al. or Bell et al. values derived here
in Table \ref{quant_table}, which give $\log (M_{dyn}/M_{\star,IMF})
\propto 0.83 \log \sigma$ and $\log (M_{dyn}/M_{\star,IMF}) \propto
0.73 \log \sigma$, respectively.

\subsection{The Thickness of the FP}\label{fp_thickness}

Having discussed the tilt of the FP in detail by examining trends
along the FP midplane, we now switch over to examining trends through
the thickness of the FP.  Figures \ref{ml_maps} and \ref{mm_maps}
showed that the high- and low-SB slices above and below the FP
midplane exhibited the same quantitative trends as the midplane, but
with zeropoint shifts in both $M_{\star,IMF}/L$ and
$M_{dyn}/M_{\star,IMF}$.  In this section, we present a detailed
exploration of the trends in both quantities through the thickness of
the plane.  

\begin{figure*}[t]
\includegraphics[width=1.0\linewidth]{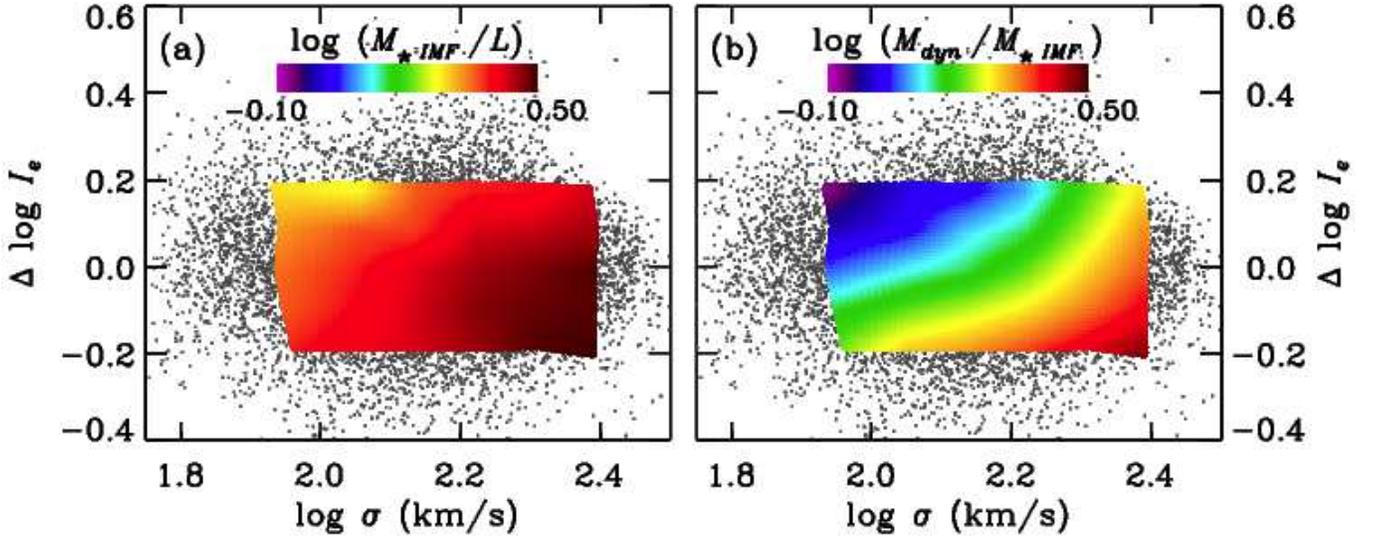}
\caption{The contribution of (a) stellar population variations and (b)
  variable dark matter/IMF to the {\it thickness} of the FP.
  Individual galaxies are plotted as points in $\Delta I_e$--$\sigma$
  space, as in Figure \ref{fp_xsection_bins}.  Color contours indicate
  $M_{\star,IMF}/L$ and $M_{dyn}/M_{\star,IMF}$ and are constructed in
  the same way as Figures \ref{ml_maps}--\ref{mm_maps}.  The color
  maps in (a) and (b) are shown on the same scale.  Both
  $M_{\star,IMF}/L$ and $M_{dyn}/M_{\star,IMF}$ vary through the
  thickness of the FP with $\Delta I_e$.  The variation in
  $M_{dyn}/M_{\star,IMF}$ is much stronger than the variation in
  $M_{\star,IMF}/L$ and therefore dominates the thickness of the FP.
}\label{ml_xsection}
\end{figure*}

\begin{figure}
\includegraphics[width=0.90\linewidth]{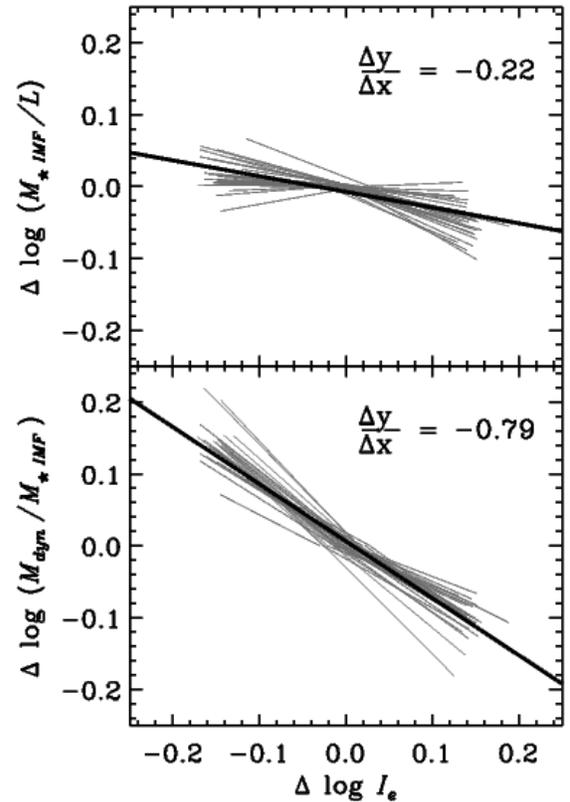}
\caption{{\it{Top:}} The change in $M_{\star,IMF}/L$ through the
  thickness of the FP.  The gray lines represent different bins in
  ($\sigma$, $R_e$), with the $M_{\star,IMF}/L$ value of the midplane
  bin subtracted off.  $M_{\star,IMF}/L$ decreases by 0.22 dex for
  every 1 dex increase in $\Delta I_e$.  {\it{Bottom:}} The change in
  $M_{dyn}/M_{\star,IMF}$ through the thickness of the FP.  Colors and
  normalization are as in the top panel.  $M_{dyn}/M_{\star,IMF}$
  decreases by 0.79 dex for every 1 dex increase in $\Delta I_e$.
  $M_{dyn}/M_{\star,IMF}$ therefore varies more than three times as
  much as $M_{\star,IMF}/L$ through the thickness of the FP.
}\label{delta_ie_ml_mm}
\end{figure}

To illustrate this, we adopt the binning strategy illustrated in
Figure \ref{fp_xsection_bins}, where bins are now defined in a 2D
cross-section through the FP.  Figure \ref{ml_xsection} maps
$M_{\star,IMF}/L$ and $M_{dyn}/M_{\star,IMF}$ across the FP
cross-section.  The same color scale is used in both panels to
facilitate comparison.  $M_{\star,IMF}/L$ and $M_{dyn}/M_{\star,IMF}$
behave similarly in this projection; both have maximum values among
high-$\sigma$, low-$I_e$ galaxies and minimum values at low $\sigma$
and high $I_e$.  Lines of constant values run diagonally from upper
right to lower left.  The main difference between these two parameters
is the range of variation.  The range of $M_{\star,IMF}/L$ values is
{\it much smaller} than the range covered by $M_{dyn}/M_{\star,IMF}$,
particularly in the $\Delta I_e$ dimension.\footnote{This projection
  of the FP averages over $R_e$.  Figures \ref{ml_maps}--\ref{mm_maps}
  showed that, while $M_{\star,IMF}/L$ is independent of $R_e$,
  $M_{dyn}/M_{\star,IMF}$ is not.  Thus the $M_{dyn}/M_{\star,IMF}$
  values mapped in Figure \ref{ml_xsection} average over a range of
  $M_{dyn}/M_{\star,IMF}$ values at each point in the FP cross-section
  and are presented for illustrative purposes only.  We have dealt
  with this variation in all quantitative presentations of the
  $M_{dyn}/M_{\star,IMF}$--$\log I_e$ $\Delta$--$\Delta$ relations
  (Table \ref{quant_table} and Figure \ref{delta_ie_ml_mm}) by
  explicitly subtracting off the dependence of $M_{dyn}/M_{\star,IMF}$
  on $R_e$.}

We have already quantified changes in $M_{\star,IMF}/L$ and
$M_{dyn}/M_{\star,IMF}$ along the plane in Figure \ref{two_tilts}.
Figure \ref{delta_ie_ml_mm} makes a similar comparison for the
variations through the thickness of the plane.  To focus on the trends
{\it through} the plane, we normalize out any trends {\it along} the
plane by subtracting mid-plane values of $M_{\star,IMF}/L$ and
$M_{dyn}/M_{\star,IMF}$ at each bin in ($\sigma$, $R_e$).  The various
($\sigma$, $R_e$) bins are represented by gray lines.  The thick black
lines show fits to the ensemble of data values, with slopes as
indicated.  Comparing these slopes shows that $M_{dyn}/M_{\star,IMF}$
varies more than three times as much through the thickness of the FP
as does $M_{\star,IMF}/L$.  Dark matter and/or IMF variations
therefore provide the dominant contribution to the thickness of the
FP.

As with the FP tilt, the relative contributions of the stellar
population and dark matter/IMF terms depends on which estimates of
$M_{\star,IMF}/L$ are used.  These quantitative differences are
summarized in the bottom section of Table \ref{quant_table}.  Using
the \citet{gallazzi05} measurements of $M_{\star,IMF}/L$, stellar
population effects contribute only 22\% of the tilt of the FP---the
remaining 78\% must be due to dark matter and/or IMF variations.  The
\citet{kauffmann03_mstar} measurements give similar results.  As we
saw with the FP tilt, using single burst models to measure
$M_{\star,IMF}/L$ through the thickness of the FP results in a larger
contribution from stellar population effects (53\%) because low values
of $M_{\star,IMF}/L$ are underestimated, while the photometry-based
measurements lead to very small contributions from stellar population
effects (2\%) because of the age-metallicity degeneracy.  Considering
the full range of models for $M_{\star,IMF}/L$, stellar populations
contribution 2--53\% of the thickness of the FP, which means that dark
matter/IMF effects contribute 47--98\%.

From Table \ref{quant_table} it is clear that, regardless of which
method is used to determine $M_{\star,IMF}$, variations in the IMF
and/or the inner dark matter fraction in galaxies are required to
explain {\it both} the tilt and the thickness of the FP.  They
contribute at least 1/2--1/3 of each effect, and it is likely that
they dominate both.

This leads to an important point.  To the extent that
$M_{\star,IMF}/L$ variations through the thickness of the FP can be
ignored (since they constitute only 22\% of the total variation using
the \citealt{gallazzi05} models), the change in $I_e$ through the
plane must be due to changes in $M_{dyn}/M_{\star,IMF}$.  However, at
a fixed point in $\sigma$ and $R_e$, $M_{dyn}$ is constant by
definition.  Thus changes in $M_{dyn}/M_{\star,IMF}$ through the plane
are primarily due to changes in $M_{\star,IMF}$ within $R_e$, which
means that {\it stellar mass surface density} ($\Sigma_{\star,IMF}$)
is changing.  In other words, the low surface brightnesses of galaxies
below the plane are due to low stellar mass surface densities, not to
the fact that their stellar populations are dim.  Galaxies do not move
below the plane by fading; they were built that way.

\begin{deluxetable*}{cccccccccccc}
\tabletypesize{\footnotesize}
\tablecaption{Fitting Formulas for Mass-to-Light Ratio
  Contributions\label{fits_table}} 
\tablehead{
\multicolumn{12}{c}{As Functions of $\sigma$, $R_e$, and $\Delta I_e$\tablenotemark{*}}
}
\startdata
$\log \phantom{.}(M_{dyn}/L_V)$           &$=$ &0.85  &$\log\sigma'$  &$+$  &0.22  &$\log R_e'$  &$-$  &1.04  &$\Delta \log I_e$  &$+$  &0.62  \\
\vspace{0.08in}  &&(0.01)  &&&(0.01)  &&&(0.01)  &&&(0.01)  \\
$\log \phantom{.}(M_{dyn}/M_{\star,IMF})$  &$=$ &0.53  &$\log\sigma'$  &$+$  &0.22  &$\log R_e'$  &$-$  &0.80  &$\Delta \log I_e$  &$+$  &0.20  \\
\vspace{0.08in}  &&(0.02)  &&&(0.01)  &&&(0.02)  &&&(0.01)  \\
$\log \phantom{.}(M_{\star,IMF}/L_V)$      &$=$ &0.31  &$\log\sigma'$  &$+$  &0.003  &$\log R_e'$  &$-$  &0.24  &$\Delta \log I_e$  &$+$  &0.42  \\
\vspace{0.08in}  &&(0.01)  &&&(0.01)  &&&(0.02)  &&&(0.01)  \\
$\log \phantom{.}(\Sigma_{\star,IMF})$     &$=$ &1.49  &$\log\sigma'$  &$-$  &1.20  &$\log R_e'$  &$+$  &0.80  &$\Delta \log I_e$  &$+$  &3.81  \\
\vspace{0.08in}  &&(0.01)  &&&(0.01)  &&&(0.02)  &&&(0.01)  \\
\hline\hline \\
\multicolumn{12}{c}{As Functions of $\sigma$, $R_e$, and $I_e$} \\
\hline \\
$\log \phantom{.}(M_{dyn}/L_V)$           &$=$ &2.00  &$\log\sigma'$  &$-$  &0.97  &$\log R_e'$  &$-$  &0.95  &$\log I_e'$  &$+$  &0.61  \\
\vspace{0.08in}  &&(0.01)  &&&(0.01)  &&&(0.01)  &&&(0.01)  \\
$\log \phantom{.}(M_{dyn}/M_{\star,IMF})$  &$=$ &1.43  &$\log\sigma'$  &$-$  &0.71  &$\log R_e'$  &$-$  &0.74  &$\log I_e'$  &$+$  &0.20  \\
\vspace{0.08in}  &&(0.03)  &&&(0.03)  &&&(0.02)  &&&(0.01)  \\
$\log \phantom{.}(M_{\star,IMF}/L_V)$      &$=$ &0.57  &$\log\sigma'$  &$-$  &0.27  &$\log R_e'$  &$-$  &0.22  &$\log I_e'$  &$+$  &0.42  \\
\vspace{0.08in}  &&(0.03)  &&&(0.02)  &&&(0.02)  &&&(0.03)  \\
$\log \phantom{.}(\Sigma_{\star,IMF})$     &$=$ &0.60  &$\log\sigma'$  &$-$  &0.27  &$\log R_e'$  &$+$  &0.74  &$\log I_e'$  &$+$  &3.82  \\
\vspace{0.08in}  &&(0.02)  &&&(0.02)  &&&(0.02)  &&&(0.01)  \\
\enddata
\tablecomments{$\sigma' \equiv \sigma / 150$ km s$^{-1}$, $R_e' \equiv
  R_e / 2.5$ kpc, and $I_e' \equiv I_e / 400 L_{\odot}$ pc$^{-2}$. \\
{*}{$\Delta \log I_e$ is defined as $\Delta \log I_e =
  \log I_e - (1.16 \log \sigma - 1.21 \log R_e + 0.55)$ (see Figure
  \ref{fp_bins}).  }}
\end{deluxetable*}

\subsection{Fitting Formulas for Mass-to-Light Ratios as Functions of
  $\sigma$, $R_e$, and $\Delta I_e$}\label{fp_fits}

In this analysis, we have quantified the variations of $M_{dyn}/L$,
$M_{dyn}/M_{\star,IMF}$, and $M_{\star,IMF}/L$ throughout 3D FP space.
These results can be summarized by fitting each of these parameters as
a function of $\sigma$, $R_e$, and $\Delta I_e$.  We assume that each
mass parameter is a first-order function of $\log \sigma$, $\log R_e$,
and $\Delta \log I_e$, then use the IDL package {\it mpfit.pro}
\citep{markwardt09} implementation of Levenberg-Marquardt minimization
to find the best such solution.  Note that because $\sigma$, $R_e$,
and $\Delta I_e$ are not orthogonal coordinates, the exponents derived
from explicitly 3D fitting do not in general exactly match those
presented from 1D fits in sections \ref{fp_tilt}--\ref{fp_thickness}.
The correct quantification of the dependence on each FP parameter must
be derived from explicitly 3D fits.

The resulting functional forms are given in the top half of Table
\ref{fits_table}, using the \citet{gallazzi05} values for
$M_{\star,IMF}/L$.  Values in parentheses under each function give the
$1\sigma$ error in each parameter of the fit, computed from the
covariance matrix of the fit.  We also include fits to the effective
stellar mass surface density, defined as $\Sigma_{\star,IMF} =
M_{\star,IMF} / (2 \pi R_e^2)$ in analogy with the effective surface
brightness.  The parameters of the $M_{\star,IMF}/L$ fit in Table
\ref{fits_table} show that $M_{\star,IMF}/L$ is independent of $R_e$,
as discussed in section \ref{fp_tilt}.  For completeness, in the
bottom section of the table we also include fits as explicit functions
of $\log I_e$, rather than using the $\Delta I_e$ parameterization.

\begin{figure*}[t]
\begin{center}
\includegraphics[width=0.9\linewidth]{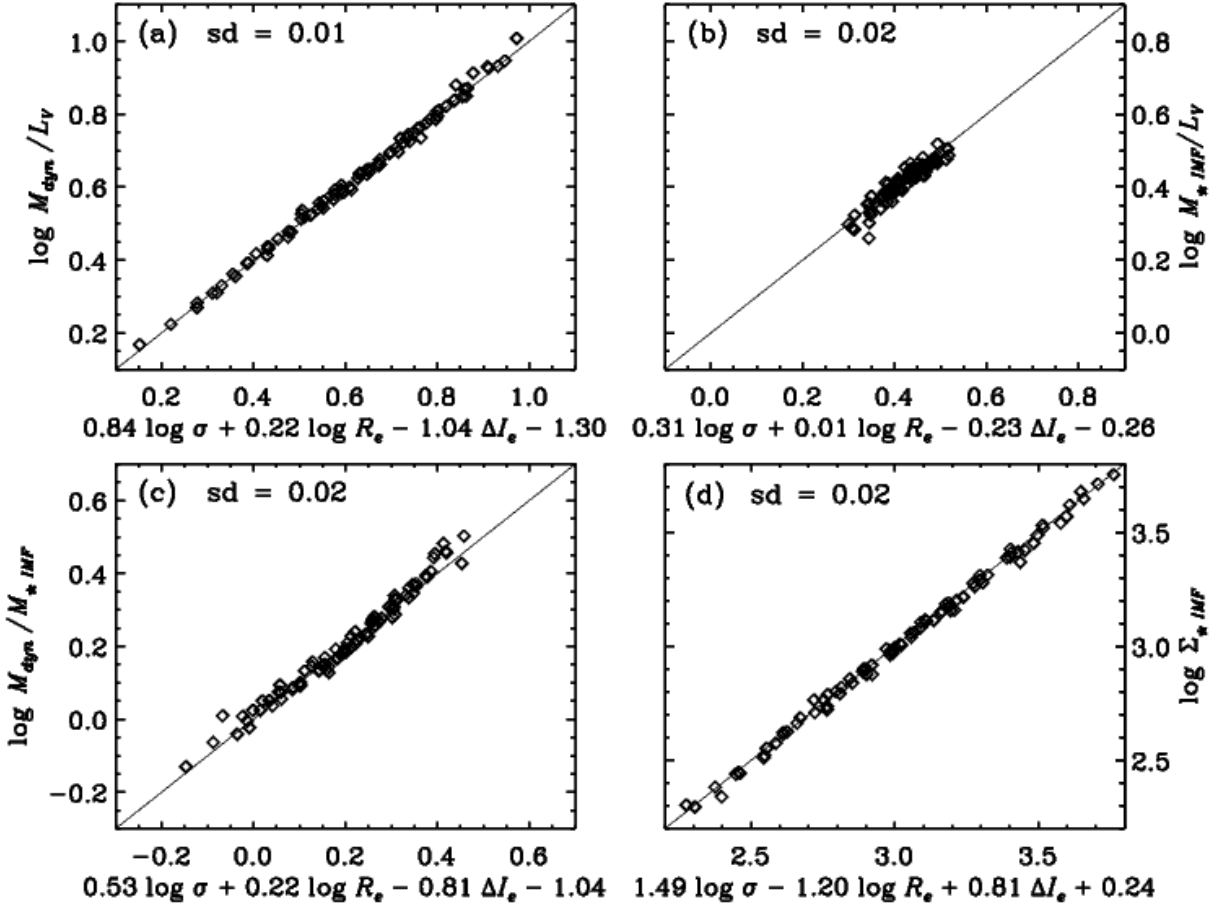}
\end{center}
\caption{The quality of fits to (a) $M_{dyn}/L_V$, (b)
  $M_{\star,IMF}/L_V$, (c) $M_{dyn}/M_{\star,IMF}$, and (d)
  $\Sigma_{\star,IMF}$ as functions of $\sigma$, $R_e$, and $\Delta
  I_e$.  In each panel, the median observed value of the parameter in
  question is plotted against the predicted value from the fits given
  in Table \ref{fits_table}.  Solid lines show the one-to-one
  relation.  The standard deviation about the one-to-one relation is
  indicated in each panel and is $\le 0.02$ dex (i.e., $\le 5$\%) for
  all four relations.  The fits given in Table \ref{fits_table} are
  therefore excellent approximations to the measured values of all
  four mass parameters.  These fits use the \citet{gallazzi05} values
  for $M_{\star,IMF}/L$.}\label{fits_figure}
\end{figure*}

Figure \ref{fits_figure} illustrates the quality of these fits.  In
each panel, we plot the median value of $M_{dyn}/L$,
$M_{dyn}/M_{\star,IMF}$, $M_{\star,IMF}/L$, or $\Sigma_{\star,IMF}$ as
measured for each bin against the value predicted by the fits in Table
\ref{fits_table}.  Panels a--c are plotted on the same scale to
facilitate comparison.  The standard deviation of the scatter about
the one-to-one relation is indicated in each panel. The fits do an
excellent job of reproducing the median observed values, with scatter
$\le 0.02$ dex (i.e., $\le 5$\% ) in all cases.  The relations can
therefore be used to estimate values of $M_{dyn}/M_{\star,IMF}$,
$M_{\star,IMF}/L$, and $\Sigma_{\star,IMF}$ for ensembles of SDSS
galaxies from the parameters $\langle \sigma \rangle$, $\langle R_e
\rangle$, and $\langle I_e \rangle$.  Note that these values are
correct in the statistical sense only, as the galaxy-to-galaxy
variation in stellar population properties within a bin cannot be
estimated from the stacking analysis presented here.

\begin{deluxetable*}{ccccccccc}
\tabletypesize{\footnotesize} 
\tablecaption{Trend Directions for Stellar
  Population and Structural Properties\label{trend_table}}
\tablewidth{0pt} 
\tablehead{ 
\colhead{} & 
\colhead{Age} &
\colhead{\protect{[Fe/H]}} & 
\colhead{\protect{[Mg/H]}} &
\colhead{\protect{[Mg/Fe]}} &
\colhead{\protect{$\Sigma_{\star,IMF}$}} &
\colhead{$M_{dyn}/M_{\star,IMF}$} &
\colhead{$M_{\star,IMF}/L$} &
\colhead{$\Delta t_f$} \\ 
\colhead{} & 
\colhead{(Gyr)} &
\colhead{(dex)} & 
\colhead{(dex)} & 
\colhead{(dex)} &
\colhead{($M_{\odot}$ pc$^{-2}$)} &
\colhead{($M_{\odot}/M_{\odot}$)} &
\colhead{($M_{\odot}/L_{\odot}$)} &
\colhead{(Gyr)} 
} 
\startdata 
$\sigma$\tablenotemark{*}
&\boldmath{$+$} &\boldmath{$+$} &\boldmath{$+$} &\boldmath{$+$}
&\boldmath{$+$} &\boldmath{$+$} &\boldmath{$+$} &\boldmath{$-$} \\ 
$R_e$\tablenotemark{\dag}
&null &null &null &null &\boldmath{$-$} &\boldmath{$+$} &null &null \\
$\Delta I_e$\tablenotemark{\ddag}
&\boldmath{$-$} &\boldmath{$+$} &\boldmath{$+$} &\boldmath{$-$}
&\boldmath{$+$} &\boldmath{$-$} &\boldmath{$-$} &\boldmath{$+$} \\
\enddata
\tablecomments{ \\
{*}{Variations with $\sigma$ along the FP, at fixed $R_e$.} \\
{\dag}{Variations with $R_e$ along the FP, at fixed
  $\sigma$.} \\
{\ddag}{Variations with $\Delta I_e$ through the
  thickness of the FP, at fixed $\sigma$ and fixed $R_e$.} }
\end{deluxetable*}

\section{Coupled Dark Matter/IMF Variations and Star Formation
  Histories Through the Thickness of the Fundamental Plane}\label{sfh} 

We have examined in detail the mass-to-light ratios of our sample of
$\sim$16,000 quiescent galaxies from the SDSS.  We have identified and
quantified the contribution of stellar population effects
($M_{\star,IMF}/L$) both along the FP and through its thickness.  By
comparing these modest variations with the much larger observed
variation in $M_{dyn}/L$, we have argued for substantial additional
variations in $M_{dyn}/M_{\star,IMF}$ due to variations in the IMF or
in the inner dark matter fractions in some galaxies.  In this
discussion, we combine this information with the stellar population
results of Paper II to evaluate three possible physical mechanisms
behind the observed $M_{dyn}/M_{\star,IMF}$ variation, with particular
focus on variations {\it through the thickness} of the FP.  

For reference, the correlations between the structural properties of
galaxies, their mass ratios, and their stellar population properties
are summarized in Table \ref{trend_table}.  This table combines the
results of this work and those of Paper II.  For each of the stellar
population properties derived in Paper II---age, [Fe/H], [Mg/H], and
[Mg/Fe]---the table indicates whether that property correlates ($+$)
or anti-correlates ($-$) with $\sigma$ (along the FP midplane at fixed
$R_e$), with $R_e$ (along the FP midplane at fixed $\sigma$), and with
$\Delta I_e$ (through the thickness of the FP).  The table also
summarizes the results of this work by indicating how
$\Sigma_{\star,IMF}$, $M_{dyn}/M_{\star,IMF}$, and $M_{\star,IMF}/L$
vary with $\sigma$, $R_e$ and $\Delta I_e$.  Lastly, the table
includes an estimated trend in the duration of star formation ($\Delta
t_{SF}$), based on the assumption that star formation timescale is
inversely correlated with [Mg/Fe] (e.g., \citealt{tinsley79,
  greggio83, worthey92, matteucci94, trager00b, thomas05}).

\begin{deluxetable*}{lccccc}
\tabletypesize{\small} 
\tablecaption{The Effects of Variable IMFs\label{imf_table}}
\tablewidth{0pt} 
\tablehead{ 
\colhead{} & 
\colhead{$M_{\star}/M_{\star,IMF}$} &
\colhead{$\Delta \log I_e$} &
\colhead{\protect{[Mg/Fe]}} &
\colhead{\protect{[Fe/H]}} & 
\colhead{\protect{[Mg/H]}} \\
\colhead{} & 
\colhead{($M_{\odot}/M_{\odot}$)} &
\colhead{($L_{\odot}$ pc$^{-2}$)} &
\colhead{(dex)} & 
\colhead{(dex)} & 
\colhead{(dex)} 
} 
\startdata 
Observed\tablenotemark{*}     &$+$ &$-$ &$+$  &$-$ &$-$ \\
More very high-mass stars     &$+$ &$-$ &$+$  &$+$ &$+$ \\
More very low-mass stars      &$+$ &$-$ &none &$-$ &$-$ \\
\enddata
\tablecomments{ \\
{*}{Assuming all variation is due to IMF differences,
  i.e., $M_{dyn}/M_{\star} = const$.}}
\end{deluxetable*}

The following sections consider three families of mechanisms for
producing $M_{dyn} / M_{\star,IMF}$ variations, focusing on variations
{\it through the thickness of the FP}.  For each of these, we examine
the implications for the stellar population properties and star
formation histories of the galaxies, and compare these to the
observations.  We find that the full set of observations is best
explained by a model that postulates the {\it premature truncation} of
star formation in galaxies with high $M_{dyn} / M_{\star,IMF}$ (and
low $I_e$)---those that lie below the FP.

\subsection{Variations in the IMF}\label{imf}

The low-$\sigma$ galaxies with high $\Delta I_e$ (upper left corner in
Figure \ref{ml_xsection}b) have log $(M_{dyn}/M_{\star,IMF}) \approx
0$, as can be read from the color-bar in that figure.  This means that
the stellar mass inferred using a Chabrier IMF can account for all of
the mass detected dynamically.  However, the rest of the galaxy bins
have log $(M_{dyn} / M_{\star,IMF}) > 0$.  If the entire discrepancy
between $M_{dyn}$ and $M_{\star,IMF}$ through the thickness of the FP
is due to IMF variations, then {\it all but the highest-$\Delta I_e$
  galaxies must have formed with IMFs that produce less light for a
  given mass of stars formed than does the Chabrier IMF}.

There are two ways this could happen: through an excess of very
low-mass stars that contribute substantially to $M_{\star}$ but
produce very little light, or through an excess of massive stars at
early times that create numerous compact remnants at late times, again
contributing substantial mass but little light.

Variations in the IMF will also change the chemical enrichment history
of a galaxy and predictions for this can be compared to the observed
abundances.  Table \ref{imf_table} summarizes the predictions for
several different models and compares them to the observations.  The
SDSS galaxies show stellar population variations through the thickness
of the FP such that galaxies with lower $\Delta I_e$ have higher
[Mg/Fe], lower [Mg/H], and lower [Fe/H] than galaxies at the same
$\sigma$ and $R_e$ with higher $\Delta I_e$ (Paper II).  These trends
are shown in the first row of Table \ref{imf_table}.

An IMF which produces more massive stars ($> 8 M_{\odot}$) at early
times (and therefore more compact remnants at late times), in addition
to having higher $M_{\star}/M_{\star,IMF}$ will produce a larger
number of SNe II relative to SNe Ia.  These galaxies should then have
higher values of [Mg/Fe].  The large number of massive stars will also
result in higher overall effective yields of both Mg and Fe, leading
to higher values of [Mg/H] and [Fe/H], as summarized in the second row
of Table \ref{imf_table}.  This combination of high [Mg/Fe] and higher
effective yields does not match the observed stellar population
variations through the thickness of the FP.

An IMF with an excess of low-mass stars could also produce high values
of $M_{\star}/M_{\star,IMF}$, such as an IMF with Salpeter slope
continuing down to low masses rather than turning over near $1
M_{\odot}$, as in the Kroupa and Chabrier IMFs.  Changing the fraction
of stars below $\sim 1 M_{\odot}$ would not affect the relative
contributions of SNe II and SNe Ia but it would result in fewer SNe
overall, thus significantly reducing the effective yields from both
types of SNe.  These lower yields are consistent with the lower values
of [Fe/H] and [Mg/H] observed in galaxies with high
$M_{dyn}/M_{\star,IMF}$, but this model does not reproduce the
increase in [Mg/Fe] seen in such galaxies (see Table \ref{imf_table}).

In general, it is difficult to simultaneously produce higher values of
[Mg/Fe] {\it and} lower yields in both [Mg/H] and [Fe/H] by adjusting
the IMF.  This is because IMF modifications that produce a larger
number of massive stars will tend to increase {\it both} [Mg/Fe] and
the effective yield.  This is not a problem when studying abundance
trends along the FP, where [Mg/Fe], [Fe/H], and [Mg/H] all increase
together with increasing $\sigma$ (e.g., Table \ref{trend_table}), but
it presents a challenge when interpreting the trends through the
thickness of the FP.

A possible way to save the IMF model is to assume that a larger
proportion of massive stars will result in increased levels of SN
feedback.  Thus, although the raw yields will be higher in galaxies
with more massive stars, a larger fraction of metal-enriched material
may be removed through intense SN feedback.  However, this involves an
additional assumption about differences in feedback between galaxies.
In what follows, we assume that the observed $M_{dyn}/M_{\star,IMF}$
variations are dominated by real differences in dark matter fraction
within $R_e$.

\subsection{Variations in Inner Dark Matter Fraction: Redistributing
  Stars and Dark Matter through Merging}\label{dm_merging}

If dark-to-stellar mass ratio is really varying through the thickness
of the FP, one of two statements must be true.  Either galaxies with
higher $M_{dyn}/M_{\star}$ are genuinely deficient in stellar mass for
their halo size, or the overall stellar mass fractions are the same
but stars are distributed differently such that the dark matter
fraction {\it within} $R_e$ is larger.\footnote{The dynamical mass
  estimator $M_{dyn}$ is predominantly sensitive to the mass inside
  $R_e$ only.}  We discuss the latter effect in this section; the
former will be discussed in the next section.

Substantial dissipation during the galaxy formation process will lead
to galaxies with centrally concentrated baryons, while less
dissipation during formation will produce more extended stellar
distributions.  Such a process has been proposed by several authors as
an explanation for the {\it tilt} of the FP.  Here, we consider
whether it might also contribute to the {\it thickness} of the FP.

The differential dissipation scenario was proposed more than a decade
ago to explain differences in the kinematic structures of galaxies
(\citealt{bender92, guzman93, ciotti96, faber97}).  More recently, it
has been modeled in the numerical simulations of several groups
(\citealt{onorbe05, kobayashi05, dekel06, robertson06}) and invoked as
an explanation for the tilt of the FP.  The amount of dissipation
during galaxy mergers is expected to be largest for low-mass galaxies,
due to their higher gas fractions (e.g., \citealt{kannappan04}), lower
surface densities, and relatively long timescales for star formation
(e.g., \citealt{noeske07b}).  With reasonable assumptions for the
dependence of dissipation on galaxy mass, \citet{robertson06} and
\citet{dekel06} are able to reproduce the tilt of the FP as a function
of $M_{\star}$ through variation in $M_{dyn}/M_{\star}$.  Recently,
\citet{covington08_thesis} has verified these results in a more
realistic cosmological context by tracing galaxy properties through
semi-analytic models of galaxy evolution, while \citet{hopkins08_fp}
have found supporting observational evidence in the light profiles of
nearby ellipticals.  Once the FP has been established, further
dissipationless merging will maintain the FP slope in isolated
elliptical galaxies \citep{capelato95, dantas03, nipoti03,
  boylan-kolchin05, robertson06}, although this is likely not true for
brightest cluster galaxies (BCGs, e.g., \citealt{lauer_bhs07}).

We have argued above that variations in $M_{dyn}/M_{\star}$ are needed
not only to reproduce the tilt of the FP, but its thickness as well.
Suppose that these variations arise from different amounts of
dissipation in the mergers that create different early type galaxies
{\it at fixed $\sigma$ and $R_e$}.  As shown in Table
\ref{trend_table}, the galaxies with lower $M_{dyn}/M_{\star,IMF}$ at
fixed ($\sigma$, $R_e$) are the galaxies with younger mean ages and
lower [Mg/Fe], consistent with their having more extended star
formation histories.  In this scenario, dissipational mergers with
their accompanying bursts of star formation would have to occur
preferentially in galaxies at later times, while early type galaxies
that had star formation truncated early would have to have experienced
a dissipationless merger.

However, this seems backwards to expectations, as galaxies at early
times were in general more gas rich.  Thus earlier mergers might be
expected to be more dissipative, not less.  In addition, the trends in
$M_{dyn}/M_{\star,IMF}$ and in stellar population properties continue
unchanged down to small galaxies with $\sigma \approx 100$ km s$^{-1}$
(see Figures \ref{ml_maps}--\ref{ml_xsection} and Figures 7--10 of
Paper II), where S0 galaxies are prevalent.  Their low values of
$M_{dyn}/M_{\star,IMF}$ require them to have experienced a
substantially dissipational merger under this scenario, after which
they would have to regrow a disk and further quench star formation in
the residual disk.  It is not clear that this merger-driven mechanism
can produce these smooth trends down to the modest galaxy masses
included in this sample.

\subsection{Variations in Inner Dark Matter Fraction: Low Efficiency
  Star Formation}\label{dm_efficiency} 

The second scenario for variations in dark matter fraction consists of
global differences in the stellar and dark matter mass fractions
within a halo.  In this scenario, a significant fraction of the
baryonic material in some haloes does not get converted into stars.
We will refer to this as a low ``conversion efficiency'' for the
baryons.  This conversion efficiency is known to vary with mass, such
that $M_{\star}/M_{halo}$ has a maximum around the characteristic
galaxy luminosity ($L^{\star}$, \citealt{benson00, marinoni02,
  vandenbosch03, zaritsky06}) and from there decreases toward lower
masses (e.g., dwarf galaxies) and toward higher masses (e.g., BCGs and
galaxy clusters).

The galaxies in our sample are on the higher-mass side of this curve
and should therefore have $M_{halo}/M_{\star}$ increasing with mass,
i.e., their conversion efficiency of baryons into stars should {\it
  decrease} with increasing mass.  The physical mechanism driving this
change is thought to involve the truncation of star formation at some
point in the evolution of the central galaxy, such that the galaxy
halo continues to accrete mass but the incoming baryons are prevented
from forming stars.  It is very interesting that
$M_{dyn}/M_{\star,IMF}$ is observed to increase with $M_{dyn}$ along
the FP (e.g., Table \ref{trend_table}).  This increase in
$M_{dyn}/M_{\star,IMF}$ could well be the same phenomenon as the
overall increase in $M_{halo}/M_{\star}$, both of which appear to
scale directly with mass. This suggests that the conversion efficiency
within $R_e$ may mimic the conversion efficiency of the parent halo as
a whole.

As we did in the previous section with dissipational merging, we can
invoke a process thought to cause $M_{dyn}/M_{\star,IMF}$ variation
along the FP to try to explain the variations through the thickness of
the FP as well.  The time at which star formation is truncated may
vary among galaxies at the same $\sigma$ and $R_e$, such that some
galaxies have their star formation shut down prematurely, while other
galaxies continue forming stars, accumulating more stellar mass, and
further enriching their ISMs for extended periods of time.  

This scenario of variable truncation times finally produces the
consistent picture we have been looking for to explain all the trends
observed through the thickness of the FP.  Galaxies at fixed $\sigma$
and $R_e$ (i.e., fixed $M_{dyn}$) which are prematurely truncated wind
up with shorter timescales for star formation (i.e., higher [Mg/Fe]),
older mean ages, lower total stars formed (lower $M_{\star}$ and lower
$\Sigma_{\star,IMF}$), and incomplete processing of metals (lower
[Fe/H] and lower [Mg/H]).  In contrast, galaxies that are not
truncated continue forming stars longer, resulting in longer star
formation timescales (lower [Mg/Fe]), younger mean ages, more total
stars (higher $M_{\star}$ and higher $\Sigma_{\star,IMF}$), and more
complete metal enrichment (higher [Fe/H] and [Mg/H]).

One possible truncation mechanism for satellites is the rapid
quenching of star formation when a satellite falls into a massive halo
and is stripped of gas (e.g., \citealt{gunn72, lea76, gisler76}).
This explanation predicts that galaxies with high
$M_{dyn}/M_{\star,IMF}$ (and therefore high [Mg/Fe]) at a given
$\sigma$ would tend to be satellites in more massive haloes today.
However, there is some evidence that this type of quenching proceeds
slowly in massive galaxies (e.g., \citealt{wolf_mhalf09}) and may not
be adequately abrupt to produce the high [Mg/Fe] seen in massive
galaxies with high $M_{dyn}/M_{\star,IMF}$.  Moreover, many of our
sample galaxies are central to their haloes---90.1\% of our sample
galaxies are identified as the most massive galaxies in their haloes
in the \citet{yang07} group catalogs---and thus satellite quenching
processes are not relevant for them.

Another possible truncation mechanism is very powerful feedback, in
which a substantial quantity of the interstellar medium (ISM) is
heated and removed from the galaxy potential well before it can form
stars.  Both starburst-driven outflows (with velocities up to $\sim
600$ km s$^{-1}$ \citealt{heckman00}) and AGN-driven outflows (which
can reach velocities in excess of $\sim 1000$ km s$^{-1}$
\citealt{trump06}) may be able to unbind gas from galaxies.
Theoretical frameworks for AGN feedback in particular have been
provided by many authors (e.g., \citealt{granato04, scannapieco04,
  di_matteo05, springel05, hopkins08_etgs}).  Real life examples of
AGN feedback could include the two massive $z \approx 3.5$ galaxiaes
discussed in \citet{nesvadba07}.  They present optical and radio
observations the strong radio jets observed in these systems and
calculate that the mass-loading of the jet material may represent
several times $10^{10} M_{\odot}$ of ISM.  Such objects may be
evidence of powerful AGN feedback in action.  SN-driven outflows may
also be able to quench star formation in early type galaxies (e.g.,
\citealt{pipino08}), although there is some evidence that the outflow
velocities in identifiable post-starburst galaxies require the
presence of an AGN \citep{tremonti07}.  

A variation on feedback-induced truncation could be supernova feedback
in smaller galaxies that then assemble into larger galaxies.  Galaxies
with high $M_{dyn}/M_{\star,IMF}$ could be galaxies that did not form
the majority of their stars {\it in situ}, but instead assembled
hierarchically at early times from many smaller galaxies.  These small
galaxies might have shallow enough potential wells that supernova
feedback can play an important role in removing baryons
\citep{white78, dekel86, white91, benson03}.  This process could shut
down star formation at earlier times and lead to the observed older
ages and higher [Mg/Fe] of their massive descendents, as compared to
galaxies at the same $\sigma$ with lower $M_{dyn}/M_{\star,IMF}$,
which may have formed more of their stars {\it in situ} in massive
haloes.  A galaxy which formed in many smaller pieces would also have
lower metallicity, both because its ISM experienced fewer generations
of stellar processing resulting in a lower effective yield, and
because SN feedback would be more effective at removing metal-enriched
gas from the galaxy.  In this scenario, the galaxies that form in
smaller pieces would also have to form at earlier times (to match the
observed age trends).

Yet another mechanism is quenching by massive haloes (e.g.,
\citealt{birnboim03, cattaneo08}), when halos pass over a critical
mass threshold ($M_{crit}$) and shock-heated gas accreting onto them
can no longer cool efficiently.  Variations through the FP could be
due to stochastic variations in the halo mass assembly history, if
galaxies with the same $\sigma$ and $R_e$ today exist in halos that
passed over $M_{crit}$ at different times, or if mass was accreted
differently onto these halos, affecting the cooling rate.  This would
mean that halo mass is not perfectly coupled to the structure of the
visible galaxy.  Also, halo-quenching scenarios need to be coupled
with some kind of maintenance mode to suppress star formation, such as
``radio-mode'' AGN activity (e.g., \citealt{croton06}).

All of these scenarios of truncated star formation are qualitatively
consistent with the higher values of [Mg/Fe], lower effective yields,
and hypothesized shorter-duration star formation histories of high
$M_{dyn}/M_{\star,IMF}$ galaxies.  Indeed, low-efficiency star
formation due to truncation is an attractive solution to
simultaneously achieving high [Mg/Fe], low effective yields, and lower
total production of stars.  Semi-analytic models of galaxy formation
may provide clues as to which, if any, of these mechanisms can produce
quantitative agreement with the observed galaxy properties.

\section{Conclusions}\label{conclusions}

In Paper II, we demonstrated that the stellar population properties,
and therefore the star formation histories, of quiescent early type
galaxies form a two-parameter family.  These map onto a cross-section
through the Fundamental Plane.  In this paper, we have explored the
associated mass-to-light variations in 3D FP space.  This has allowed
us to map out the various components of mass-to-light variation both
along the FP and through its thickness, with the following results:

\renewcommand{\labelenumii}{\alph{enumii})}
\begin{enumerate}
\item{We map $M_{dyn}/L$, $M_{dyn}/M_{\star,IMF}$, and
  $M_{\star,IMF}/L$ throughout 3D Fundamental Plane space.  These
  quantities can be well-approximated by first-order functions (i.e.,
  hyperplane fits) of $\log \sigma$, $\log R_e$, and $\Delta \log
  I_e$.  These fits are provided in Table \ref{fits_table}.}
\item{We confirm that the variation in $M_{\star,IMF}/L$ due to
  stellar population variations cannot account for the tilt of the FP,
  as demonstrated by many previous authors.  Variations in the dark
  matter fraction within $R_e$ (i.e., $M_{dyn}/M_{\star}$) or in the
  IMF (i.e., $M_{\star}/M_{\star,IMF}$) are required to explain the
  tilt of the FP.  Using the $M_{\star,IMF}$ estimates of
  \citet{gallazzi05}, stellar population effects contribute only 1/3
  of the tilt of the FP, while dark matter/IMF effects contribute the
  remaining 2/3.}
\item{The stellar population and the dark matter/IMF contributions to
  the FP tilt rotate the plane around different axes in the 3D space,
  with $M_{\star,IMF}/L \propto f(\sigma)$, while
  $M_{dyn}/M_{\star,IMF} \propto g(M_{dyn})$. }
\item{Although Paper II showed that stellar populations vary through
  the thickness of the FP (in agreement with \citealt{forbes98},
  \citealt{wuyts04}, and \citealt{gargiulo09}), the associated
  variations in $M_{\star,IMF}/L$ alone {\it cannot account for the
    thickness of the FP}.  Variations in the dark matter contribution
  within $R_e$ (i.e., $M_{dyn}/M_{\star}$) or in the IMF (i.e.,
  $M_{\star}/M_{\star,IMF}$) are required to explain the thickness of
  the FP (as well as the tilt).  Using the $M_{\star,IMF}$ estimates
  of \citet{gallazzi05}, stellar population effects contribute only
  22\% of the thickness of the FP, while dark matter/IMF effects
  contribute the remaining 78\%.}
\item{Because stellar population effects contribute only $\sim 22$\%
  of the thickness of the FP, the observed $M_{dyn}/L$ variations
  through the plane are dominated by variations in $\Sigma_{\star,IMF}$.
  This means that galaxies do not move below the plane by fading; they
  either form initially with lower $\Sigma_{\star,IMF}$ or move there due
  to {\it structural} changes in the stellar mass distribution.  }
\item{Combining the results of this work with the results of Paper II,
  we find that variations in the dark matter/IMF contribution through
  the thickness of the FP are associated with differences in galaxy
  star formation histories.  The correlation is such that galaxies
  with higher $M_{dyn}/M_{\star,IMF}$ also have higher [Mg/Fe], older
  ages, and lower metallicities than their
  lower-$M_{dyn}/M_{\star,IMF}$ counterparts at the same $\sigma$,
  which may indicate that the higher-$M_{dyn}/M_{\star,IMF}$ galaxies
  have experienced shorter duration star formation.  Possible physical
  mechanisms for producing high-$M_{dyn}/M_{\star,IMF}$ and the
  associated stellar population variations include:
\begin{enumerate}
\item{A top-heavy IMF in some galaxies, which produces more SNe II
  relative to SNe Ia and therefore higher [Mg/Fe].  However, this
  scenario does not simultaneously produce the observed lower [Fe/H]
  and [Mg/H] in these galaxies.  Invoking IMF variations to explain
  the observed $M_{dyn}/M_{\star,IMF}$ variations would require an
  additional physical mechanism, such as enhanced SN feedback that
  removes a substantial fraction of the metal-enriched material.}
\item{Higher dark matter fraction within $R_e$ due to low-dissipation
  merging in galaxies that have already quenched their star formation.
  This scenario is plausible, but the predicted trend with stellar
  population age seems backwards to expectations since early-forming
  galaxies are naturally more gas-rich, not less.}
\item{Low conversion efficiency of baryons into stars through the
  premature truncation of star formation.  There are various possible
  mechanisms for this trunction, including ram-pressure stripping in
  satellite galaxies, powerful AGN feedback, massive halo quenching,
  or low-mass progenitor galaxies that are strongly affected by
  supernova feedback (as compared to galaxies that form most of their
  stars {\it in situ} in deep potential wells).  This truncation
  scenario is appealing because it simulatenously matches the lower
  stellar masses, older ages, higher [Mg/Fe], lower [Fe/H], and lower
  [Mg/H] observed in the high-$M_{dyn}/M_{\star,IMF}$ galaxies at
  fixed $\sigma$.}
\end{enumerate}
}
\item{There is an intriguing parallel between the known increase of
  $M_{halo}/M_{\star}$ in large haloes and the increase in
  $M_{dyn}/M_{\star,IMF}$ along the FP observed in this galaxy sample.
  This suggests that the conversion efficiency of baryons into stars
  inside $R_e$ mimics the overall conversion efficiency for the parent
  halo.  The truncation of star formation in massive galaxies is
  widely accepted as the explanation for the increase in
  $M_{halo}/M_{\star}$ with galaxy mass, although the truncation {\it
    mechanism} is debated.  It may be that this same truncation
  mechanism also produces differences in conversion efficiency {\it
    inside} $R_e$, and that this effect drives the observed variation
  in $M_{dyn}/M_{\star,IMF}$ through the thickness of the FP as well.}
\end{enumerate}

These results have focused on quiescent galaxies only, excluding early
type galaxies with ongoing star formation, Seyfert, or LINER activity.
Seyfert and LINER hosts are known to have systematically younger
stellar population ages than their quiescent counterparts
\citep{kauffmann03_agn, graves07, schawinski07}, but further analysis
is required to determine whether they follow the other stellar
population and $M/L$ trends explored here for quiescent galaxies.

Taken as an ensemble, the results presented in this series of papers
illustrate a connection between the star formation histories and the
present-day mass structures of early type galaxies.  As such, they
represent a stringent test for models of galaxy formation and a
challenge for the next generation of semi-analytic models.  They also
highlight the emerging importance of abundances in testing model star
formation histories.  Paper IV makes extensive use of this abundance
information to expand on and quantify the premature truncation model
proposed here.

\acknowledgements

The authors would like to thank Renbin Yan for providing the emission
line measurements used to identify the sample of early-type galaxies
used here.  This work was supported by National Science Foundation
grant AST 05-07483.  G. G. is supported by a fellowship from the
Miller Institute for Basic Research in Science.  G. G. also
acknowledges support from the ARCS Foundation and from a UCSC
Dissertation Year Fellowship

Funding for the creation and distribution of the SDSS Archive has been
provided by the Alred P. Sloan Foundation, the Participating
Institutions, the National Aeronautics and Space Administration, the
National Science Foundation, the US Department of Energy, the Japanese
Monbukagakusho, and the Max-Planck Society. The SDSS Web site is
http://www.sdss.org/.

The SDSS is managed by the Astrophysical Research Consortium (ARC) for
the Participating Institutions. The Participating Institutions are the
University of Chicago, Fermilab, the Institute for Advanced Study, the
Japan Participation Group, the Johns Hopkins University, the Korean
Scientist Group, Los Alamos National Laboratory, the
Max-Planck-Institute for Astronomy (MPIA), the Max-Planck-Institute
for Astrophysics (MPA), New Mexico State University, University of
Pittsburgh, University of Portsmouth, Princeton University, the United
States Naval Observatory, and the University of Washington.

\bibliographystyle{apj}
\bibliography{apj-jour,myrefs}

\appendix

\section{Correcting for Low-level H$\beta$ Emission Infill}\label{oIII}

\begin{figure}[t]
\begin{center}
\includegraphics[width=0.9\linewidth]{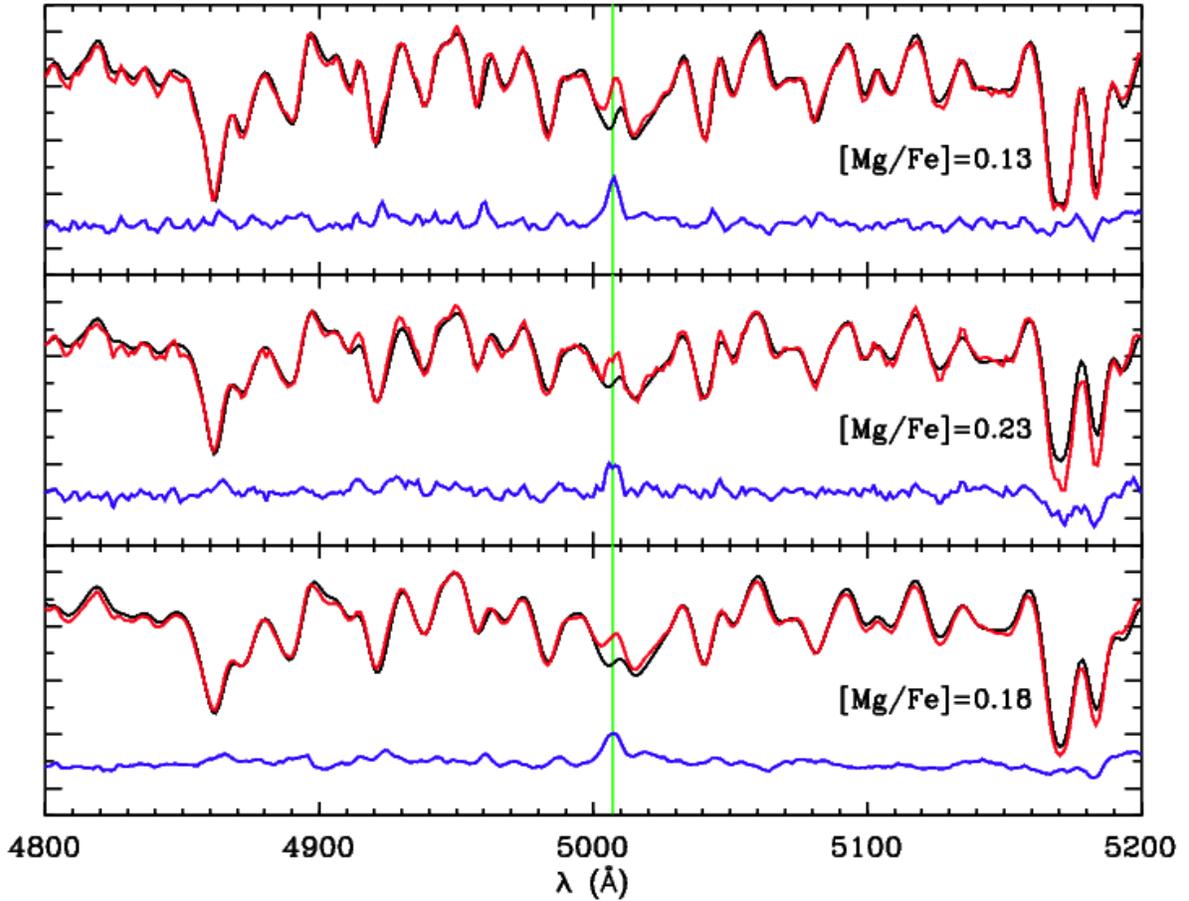}
\end{center}
\caption{Three examples of weak [O\textsc{iii}]$\lambda$5007 emission
  in the stacked spectra.  In each panel, the thick black line shows
  the data, the thin gray line shows the BC03 model with the closest
  match in age and [Fe/H], and the lower dark gray line shows the
  difference between the data and the model.  The vertical line
  indicates the location of the [O\textsc{iii}]$\lambda$5007 emission
  line, which is weak but clearly detected in all cases.  The data
  have, from top to bottom, $\log \sigma = 2.06$, $\log \sigma =
  2.14$, and $\log \sigma = 2.23$, with the models smoothed to match
  the combined SDSS spectral resolution and intrinsic velocity
  dispersion of the data.  The BC03 models have a fixed abundance
  pattern, while many of the sample galaxies have substantially
  super-solar [Mg/Fe] (as labelled).  This causes a mismatch between
  the data and models around the 5190~{\AA} Mg absorption feature but
  does not seriously bias the fit to most of the spectrum, which is
  dominated by Fe absorption features.  }\label{oIII_measurement}
\end{figure}

The galaxies presented here are selected to contain no ionized gas
emission by requiring that the strong emission lines at H$\alpha$ and
[O\textsc{ii}]$\lambda$3727 be undetected at the $2\sigma$ level.
This results in a sample of quiescent galaxies that lack both
significant ongoing star formation and AGN/LINER activity.

The individual SDSS spectra have relatively low $S/N$.  Our sample
therefore contains galaxies with weak emission below the detection
threshold in individual galaxies.  Weak H$\beta$ emission will
partially fill in the H$\beta$ absorption feature used to determine
galaxy ages, causing the stellar population modelling to return ages
that are too old.  Even small errors in H$\beta$ can result in
significant overestimates of the galaxy ages; at 10 Gyr, 0.1 {\AA} in
H$\beta$ corresponds to an age difference of roughly 1 Gyr
\citep{schiavon07}.  However, if weak emission can be detected in the
high-$S/N$ stacked spectra, a correction for this effect can be
applied to the measured values of H$\beta$, resulting in more accurate
age measurements.

\begin{figure}[t]
\begin{center}
\includegraphics[width=0.8\linewidth]{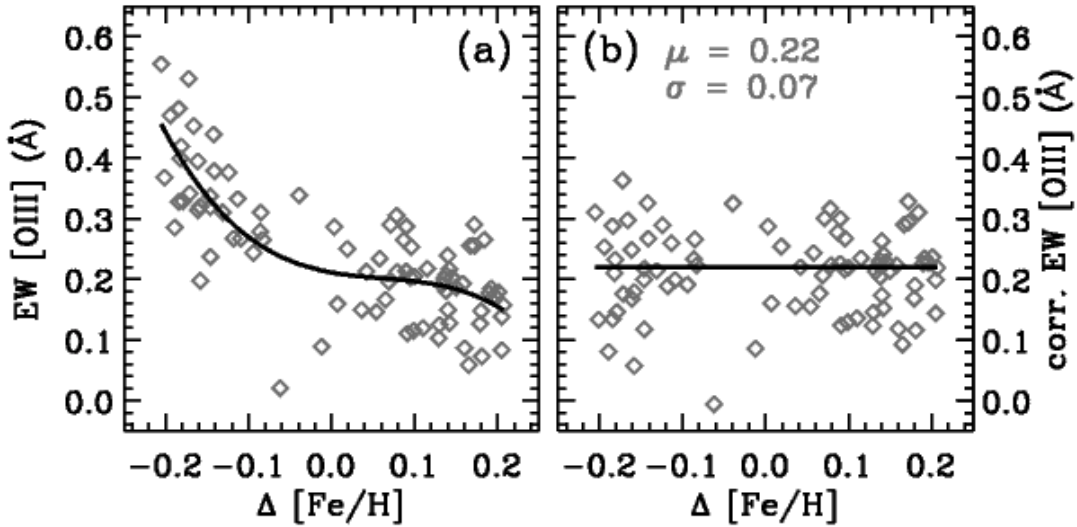}
\end{center}
\caption{(a) The EW of [O\textsc{iii}]$\lambda$5007 as a function of
  the offset in [Fe/H] between the measured [Fe/H] of the data and the
  closest-matching BC03 model.  Mismatches in [Fe/H] between the data
  and the model cause biases in the estimates of [O\textsc{iii}]. We
  use a third-degree polynomial fit (solid line) to fit the trend in
  [O\textsc{iii}] with $\Delta$[Fe/H] and correct the [O\textsc{iii}]
  measurements to the values that they would have if $\Delta$[Fe/H]$ =
  0$ for the entire sample (b).  The mean and standard deviation of
  the corrected [O\textsc{iii}] EW measurements are given.
}\label{dfeh_oIII}
\end{figure}

We do this by measuring the equivalent width (EW) of emission in the
[O\textsc{iii}]$\lambda$5007 line and using the relation
\begin{equation}\label{hbeta_equation}
\Delta\mbox{H}\beta = 0.7 \mbox{EW}([\mbox{O\textsc{iii}}] \lambda 5007)
\end{equation}
suggested by \citet{trager00a}.  Using the sample of \citet{graves07},
we have confirmed that this relation is a good approximation for the
line ratios of LINERs, which likely make up the majority of weak
emission-line objects in our sample (\citealt{yan06}, see also Figure
7b of that work).

To measure EW[O\textsc{iii}], we restack all of the composite spectra
without smoothing the spectra to $\sigma = 300$ km s$^{-1}$.  A weak,
narrow emission feature should be easier to detect if it is not
smoothed up to the maximum velocity dispersion of the sample.  In
practice, the width of [O\textsc{iii}] appears to track the stellar
$\sigma$ closely, such that the emission in high-$\sigma$ galaxies is
substantially broader than the emission in low-$\sigma$ galaxies,
although the integrated EW is similar.

For each stacked spectrum, we identify the BC03 model with the closest
match to the age and [Fe/H] values measured in our stellar population
analysis.  The BC03 models sample parameter space fairly densely in
age, but only sparsely in [Fe/H], so that the age mismatch between
data and model is less than 0.02 dex, while the mismatch in [Fe/H] can
be as much as 0.2 dex.  Three examples with different $\sigma$ values
are shown in Figure \ref{oIII_measurement}.  We divide out the
continua of the data and the model, smooth the model to match the
combined SDSS resolution and intrinsic $\sigma$ of the stacked
spectrum, and compare the data (red line) and model (black line) in
the 4800--5200~{\AA} range.

The models are an excellent match to the data except in two locations:
[O\textsc{iii}] emission, though weak, is clearly visible at
5007~{\AA} in the data but is not included in the models, and the data
show stronger absorption in the Mg feature at 5190~{\AA} than is
predicted in the models.  The latter effect is due to the fact that
the BC03 models have a fixed solar abundance pattern, while early type
galaxies typically have super-solar [Mg/Fe].  The larger the measured
value of [Mg/Fe], the bigger the mismatch between data and model at
5190~{\AA}.  The mismatch in the abundance pattern does not seriously
bias the fit to the rest of the spectrum, which is dominated by Fe
absorption features

To measure [O\textsc{iii}], we compute the difference spectrum between
the data and the model (blue line) and integrate the residual
[O\textsc{iii}] flux, using the bandpass and continuum definitions of
\citet{yan06}.  The detection of [O\textsc{iii}] emission implies that
H$\beta$ emission is also present, filling in the stellar H$\beta$
absorption line.  This H$\beta$ infill is not visible because the
chosen BC03 model has been matched to the mean age of the {\it
  uncorrected} stacked spectrum.

As mentioned above, the BC03 models are only computed for a limited
set of [Fe/H] values and are therefore not always well-matched to the
data.  Metallicity offsets between the data and the model can bias the
measurement of EW([O\textsc{iii}]).  Figure \ref{dfeh_oIII}a shows the
measured value of EW([O\textsc{iii}]) as a function of the offset
between [Fe/H] as measured in the data and the closest-matched BC03
model.  Compared to galaxies that are well-matched to the models
($\Delta$[Fe/H] $\approx 0$), galaxies for which the best-fitting
model is at higher metallicity ($\Delta$[Fe/H] $< 0$) or lower
metallicity ($\Delta$[Fe/H] $> 0$) produce EW([O\textsc{iii}])
measurements that are significantly overestimated or underestimated,
respectively.  

To correct our EW([O\textsc{iii}]) measurements to the value they
would have if a perfectly-matched model were available, we fit a
third-degree polynomial to the data in Figure \ref{dfeh_oIII}a and
subtract out the trend with $\Delta$[Fe/H] to produce corrected
EW([O\textsc{iii}]) values (Figure \ref{dfeh_oIII}b).  The corrected
EW([O\textsc{iii}]) measurements have a roughly Gaussian distribution
with mean $\mu = 0.22$ and standard deviation $\sigma = 0.07$.  There
is no correlation between the corrected EW([O\textsc{iii}]) and galaxy
$\sigma$, $R_e$, $I_e$, $\Delta I_e$, age, [Fe/H], [Mg/Fe], or
mismatch between the measured age and the closest BC03 model age.

We use the corrected EW([O\textsc{iii}]) measurements to estimate a
correction to H$\beta$ using equation \ref{hbeta_equation}.  These
corrections are subtracted from the measured H$\beta$ line strengths.
We then rerun the stellar population analysis on the corrected data.
Correcting for emission infill produces ages that are $\sim 0.12$ dex
lower than the uncorrected values, while [Fe/H] values increase by
$\sim 0.06$ dex.  

These corrected ages are of course different from the original ages
used to fit the BC03 models and measure [O\textsc{iii}] emission.  To
check that the EW([O\textsc{iii}]) values are reasonable, we iterate
the process, using the corrected ages to fit BC03 models to the
stacked spectra and remeasure EW([O\textsc{iii}]).  The resulting
values of EW([O\textsc{iii}]) are typically 0.02~{\AA} smaller (with
standard deviation 0.02~{\AA}) than the original estimate, which
translates into H$\beta$ differences of $< 0.015$.  These differences
are negligible.  

\begin{figure}[t]
\includegraphics[width=1.0\linewidth]{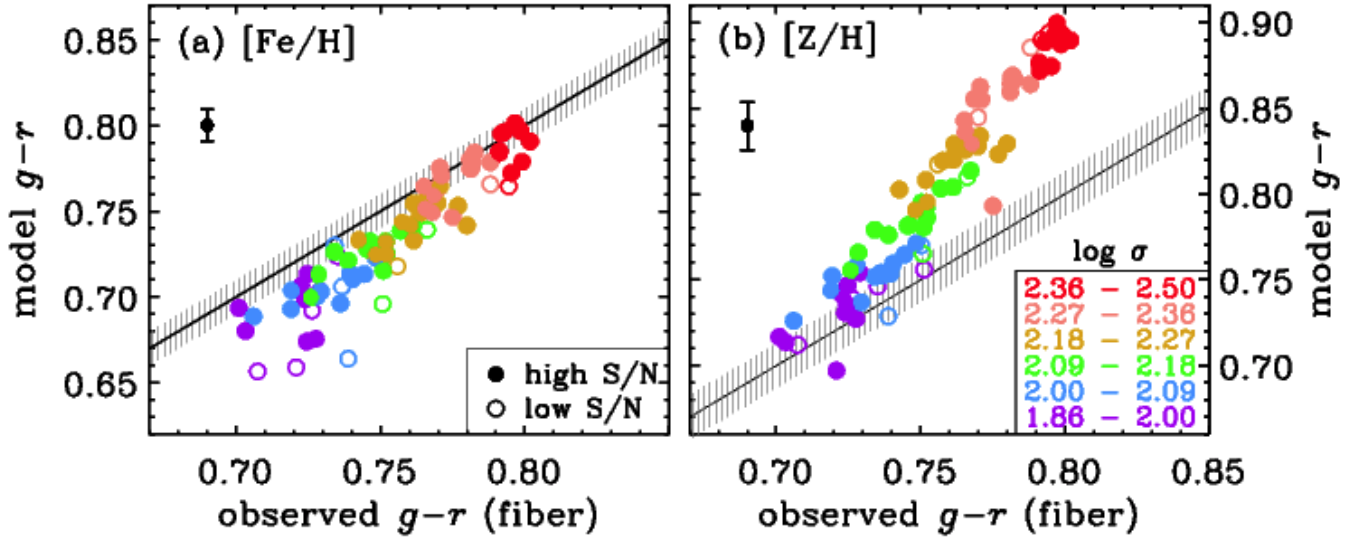}
\caption{(a) A comparison of the $g-r$ colors derived from stellar
  population modelling with the observed $g-r$ colors as measured from
  photometry matched to the $3''$ SDSS spectral fiber aperture.  Model
  $g-r$ colors are derived by matching the age and [Fe/H] measured for
  the stacked galaxy spectra to BC03 models, as illustrated in Figure
  \ref{hyperplane}b.  The line shows a one-to-one relation, with
  shading indicating the typical $1\sigma$ scatter expected due to
  absorption line measurement errors.  Data with substantially larger
  errors are shown as open circles.  For high-$\sigma$ galaxies, the
  model colors are an excellent match to the observed colors, with
  observed colors showing $\leq 0.02$ mag of internal reddening that
  is not included in the models.  The discrepancy is larger for
  low-$\sigma$ galaxies, suggesting that single burst models
  systematically underestimate the mean stellar population ages for
  younger galaxies (see text for details).  (b) The same as (a), but
  with models derived by matching the age and [Z/H] measured in the
  stacked spectra.  Matching [Z/H] instead of [Fe/H] results in model
  colors that are substantially too red.  }\label{bc03_colors}
\end{figure}

\section{Comparing Observed and Modeled Galaxy
  Colors}\label{compare_colors} 

By comparing the color estimates from Figure \ref{hyperplane}b to the
observed galaxy colors, we can check that this method of matching BC03
models to the data gives reasonable estimates of the stellar
population properties of the sample galaxies.  The spectroscopic data
sample only the region of the galaxy covered by the SDSS spectral
fibers, but many galaxies show significant color gradients.  We
therefore compare the predicted model $g-r$ colors from the
spectroscopic data to the observed galaxy colors {\it as measured from
  SDSS photometry matched to the $3''$ SDSS spectral fiber aperture.}
The fiber magnitudes are downloaded from the SDSS Catalog Archive
Server\footnote{http://cas.sdss.org/dr6/en/} and K-corrected to $z=0$.
We use the median value of the $g-r$ fiber color from all the
constituent galaxies in the corresponding bin to determine the
``observed'' $g-r$ fiber color of each stacked spectrum.

Figure \ref{bc03_colors}a shows the comparison between the observed
$g-r$ fiber colors and those derived from the spectroscopic single
burst models using the curves in Figure \ref{hyperplane}b.  The
various colors indicate the values of $\sigma$ in each bin, as
labelled.  The solid line shows a one-to-one relation between observed
and modeled colors, with the shaded region showing the typical
1$\sigma$ error on the model colors as propagated through the stellar
population modeling.  Bins whose stacked spectra have considerably
lower $S/N$ (i.e., errors in the model colors more than twice as
large) than the typical bins are shown as open circles.  For the
high-$\sigma$ galaxies, the model colors are an excellent match to the
observed colors.  The observed colors are offset not more than 0.02
dex redder than the model colors derived for these galaxies, which may
indicate low-level internal dust reddening within the sample galaxies.
No reddening has been applied to the models.

At lower values of $\sigma$, the discrepancy between observed and
modeled colors increases; the galaxies are observed to be 0.03--0.04
dex redder than the model colors.  This systematic trend, wherein the
models predict colors for the bluest galaxies that are systematically
too blue, highlights the probable bias inherent in using single burst
models.  The Balmer absorption lines are highly sensitive to a small
``frosting'' of young stars---more so than the total galaxies
colors--and will therefore give age estimates that are systematically
younger than models with composite star formation histories.  These
will then lead to predicted colors that are too blue and, as we show
in section \ref{mstar_l}, to $M_{\star,IMF}/L$ values that are too
low.

Given these considerations, the colors derived from the modelling
illustrated in Figure \ref{hyperplane} appear to do an excellent job
of matching the observed galaxy colors.  Furthermore, they reveal the
effect of an expected systematic bias in the modeling process.  

A further important issue is also clarified by matching the model and
observed colors.  The BC03 models are constrained to use the solar
abundance ratios at all metallicities.  However, massive early type
galaxies are known to have super-solar enhancements of a number of
elements, including Mg (e.g., \citealt{worthey92}) and C and N (e.g.,
\citealt{graves07, smith09_abundances}).  Since a mismatch between the
galaxy abundance patterns and the model abundance patterns is
therefore inevitable, it is not obvious {\it a priori} whether one
should attempt to match [Fe/H] between the data and the models, or
whether it would be more appropriate to match total metallicity
([Z/H]).

Figure \ref{bc03_colors}a was constructed by matching [Fe/H] between
the data and the models.  Figure \ref{bc03_colors}b shows results when
the match is instead based on [Z/H].  For the observed galaxies, [Z/H]
is calculated based on the assumption that the $\alpha$-elements O,
Ne, Na, Si, S, and Ti all track Mg, that the Fe-peak elements Cr, Mn,
Co, Ni, Cu, and Zn all track Fe, and that Mg, Fe, C, N, and Ca have
the abundances measured using EZ\_Ages.  All other elements are
assumed to have solar abundances (see \citealt{graves07}, section
5.3).

Matching models to the data on the basis of [Z/H] clearly does not do
as well as matching based on [Fe/H].  The modeled colors are
substantially redder than the observed colors for the high-$\sigma$
galaxies.  

The good match achieved between the observed and modeled colors using
[Fe/H] motivates us to use the same models to measure
$M_{\star,IMF}/L$.  We do this in section \ref{single_burst_ml},
keeping in mind the caveat that single burst models tend to produce
colors that are too blue for the bluest galaxies and will similarly
underestimate $M_{\star,IMF}/L$ for the lowest $M_{\star,IMF}/L$
galaxies.

\end{document}